\documentclass[aps,prb,superscriptaddress,twocolumn,showpacs,notitlepage]{revtex4-1}

\usepackage{amsmath} \usepackage{amsfonts} \usepackage{amssymb}
\usepackage{hyperref} \usepackage{graphicx} \usepackage{color}
\usepackage{xcolor}
\usepackage[mathscr]{euscript}
\usepackage{rotating}  %
\usepackage{mathtools}
\usepackage{pbox}

 \newcommand{\ud}[1]{{#1^{\dagger}}}
\newcommand{\bra}[1]{\left\langle #1\right|}
\newcommand{\ket}[1]{\left| #1\right\rangle}

\newcommand{\rhoel}[1]{\ket{#1}\bra{#1}}

\newenvironment{proposition}{\paragraph*{\textbf{Proposition:}}}{\hfill$\square$\\}

\begin{document} 
\title{Exciting with Quantum Light. I.~Exciting an harmonic oscillator.}

\author{J.~C.~L\'{o}pez~Carre\~{n}o} 
\affiliation{Departamento
de F\'isica Te\'orica de la Materia Condensada, Universidad
Aut\'onoma de Madrid, 28049 Madrid, Spain}

\author{F.~P.~Laussy} 
\affiliation{Russian Quantum Center, Novaya 100, 143025 Skolkovo, Moscow Region, Russia}
\affiliation{Departamento de F\'isica Te\'orica de la Materia Condensada, Universidad Aut\'onoma de Madrid, 28049 Madrid, Spain} 

\begin{abstract} 
  We start a series of studies of the excitation of an optical target
  by quantum light. In this first part, we introduce the problematic
  and address the first case of interest, that of exciting the quantum
  harmonic oscillator, corresponding to, e.g., a single-mode passive
  cavity or a non-interacting bosonic field. We introduce a mapping of
  the Hilbert space that allows to chart usefully the accessible
  regions. We then consider the quantum excitation from single photon
  sources in the form of a two-level system under various regimes of
  (classical) pumping: incoherent, coherent and in the Mollow triplet
  regime.  We close this first opus with an overview of the material
  to be covered in the subsequent papers.
\end{abstract}

\date{\today} 
\maketitle 

\section{Introduction} 

Photonics\cite{saleh_book07a} has been highly successful in the
engineering of quantum sources since the proof-of-principle production
of quantum light in the mid-seventies.\cite{clauser74b,kimble77a}
Befitting its status of the elementary brick of the light field, the
single photon was the first type of genuinely non-classical type of
light and gave rise to the concept of a single photon source
(SPS).\cite{michler00a,shields07a,eisaman11a} Nowadays, SPS abound
with ever increasing figures of
merit\cite{kuhlmann15a,portalupi15a,muller15a,ding16a,somaschi16a} and
are even commercially available. The motivations for SPS are
many,\cite{chunnilall14a} from metrology\cite{ware04a,taylor16a} to
input for quantum information
processing,\cite{beveratos02a,kiraz04a,tamma16a} passing by
bio-technology,\cite{lambert13a,tinsley16a} sensing and
detecting,\cite{hadfield09a,reithmaier13a} etc. In most cases, the
technology is still under development and quantum light is not yet
deployed on the market. For this reason, the focus is still largely on
the source itself rather than on its direct use as part of a
technological component. There is however an increasing interest in
using quantum light for actual applications. For instance, there have
been recently converging propositions from independent groups to use
quantum light for
spectroscopy.\cite{kira06a,assmann11b,assmann15a,mukamel15a,lopezcarreno15a}

Thanks to the theory of frequency-resolved photon
correlations,\cite{delvalle12a} it was shown how quantum sources can
be greatly tuned in their characteristics by selecting light in astute
frequency windows, revealed by the theory in the form of so-called
``two-photon correlations
spectra''.\cite{delvalle13a,gonzaleztudela13a} This can be used to
identify and exploit unsuspected types of quantum
correlations~\cite{sanchezmunoz14b} and/or enhance them by learned
combination of timescales and frequencies,\cite{gonzaleztudela15a}
merely by spectral filtering of a quantum emitter.\cite{kamide15a}
Beyond optimizing parameters to find the best compromises for sought
applications, the theory also reveals that interesting quantum signal
is typically not found at the expected spectral locations such as
peaks, that correspond to classical-like de-excitation of the emitter
between real states at the one-photon level. Instead, strong quantum
correlations arise from, e.g., two-photon de-excitation involving
intermediate virtual states.  Such channels of de-excitation were
termed ``\emph{leapfrog
  processes}''.\cite{delvalle13a,gonzaleztudela13a} They are emitted
in unremarkable frequency windows at the photo-luminescence level. At
the quantum-optical level, however, they are the quintessence of
quantum emission. This picture has been confirmed experimentally by
A.~M\"uller's group.\cite{peiris15a} A closer look at, and proper
selection of, the photons emitted by quantum sources thus appears
fundamental for state-of-the-art quantum applications, their
correlations being otherwise averaged over often competing types. This
viewpoint of frequency-resolved photon correlations therefore poses in
a new light the problem of the excitation of optical targets with such
inside-knowledge of the features of the sources. This revives a
question put to close scrutiny by Gardiner~\cite{gardiner93a} and
Carmichael~\cite{carmichael93b} in 1993 following the emergence of
sources of squeezed light,\cite{walls83a} which appeared sufficiently
more elaborate as compared to SPS to warrant a direct investigation of
how they would affect optical targets as compared to conventional
types of excitation (classical fields, possibly stochastic). Resonance
fluorescence in the squeezed vacuum has in fact been just recently
reported.\cite{toyli16a} Gardiner and Carmichael's (independent)
treatment of the problem of quantum excitation in two consecutives
Letters in the Physical Review~\cite{gardiner93a,carmichael93b}
achieved the setting up of a formalism---named the ``\emph{cascaded
  formalism}'' by Carmichael---that allows to excite a system (which
we will call the ``target'') by an other (the ``source'') without
back-action from the target to the source. This permits to think
separately of the quantum source, which properties can be first
studied (through the two-photon spectrum, for instance) and then
directed onto a target.  For historical accuracy, let us mention that
the problem was first contemplated by Kolobov and
Sokolov~\cite{kolobov87a} who tackled it by providing all the
correlations of the exciting quantum field.  This was recognized as an
overkill by Gardiner and Carmichael~\cite{gardiner93a,carmichael93b}
(Gardiner had made prior attempts along these lines). They proposed
instead to model the quantum source dynamics as well as the response
of the target, even if only the latter is of interest. From our point
of view, such a treatment is essential since the frequency
correlations of the source are too complicated to be treated otherwise
than fully and explicitly by solving the complete problem. Also, such
correlations are dynamical in character, and cannot be well
approximated by quantum states as initial conditions. Instead, they
must be dealt with through the full apparatus of dissipative quantum
optics, feeding the target with the complete treatment of virtual
states and other types of strongly correlated quantum input. The
cascaded formalism is therefore particularly apposite for exciting
optical systems with the knowledge of the two-photon correlations
spectrum of an emitter.

Despite the conceptual importance of quantum excitation, there have
been a moderate follow-up of this cascaded formalism, which we believe
is a deep and far-reaching contribution to the problem of light-matter
interaction.  Even though it became textbook material (see the last
chapter from one of the pioneering authors~\cite{gardiner_book00a})
and generated a sizable amount of citations, few texts do actually
fully exploit the idea. Gardiner and Parkins (the formalism is
sometimes also named after these two authors) undertook a more
thorough analysis of various types of non-standard statistics of the
source~\cite{gardiner94a} and Cirac \emph{et al.}~used it to describe
perfect transmission in their proposal for a quantum
network,\cite{cirac97b} but overall, the core of the literature using
the formalism focuses on specialized particular cases, such as driving
with squeezed light.\cite{smyth99a,messikh00a} Typically, the
discussion is then held at the level of correlations from a quantum
state (namely a squeezed state), as opposed to dynamical correlations
from a quantum source.  The other studies, already evoked, turned to
approximate or indirect approaches, quite similar to the earlier
attempts before the cascaded formalism was set up. The reasons for
this is certainly a mix between convenience of using well-known and
established formalisms and the as-yet unclear advantages of the
alternative one.

In this series of texts, we make an extensive study of exciting with
quantum light (with no feedback of the target to the source), and show
that some new features of light-matter interaction emerge, making the
overall problem in need of scrupulous attention. The formalism itself
needs little further development and we will mainly adapt it to new
cases but with the additional knowledge provided by frequency-resolved
correlations (we will extend the formalism in the following papers to
consider sequences of cascades and multiple sources). We briefly
introduce its core machinery in a self-contained way in
Section~\ref{sec:marnov24124242CET2015} as a convenience for the
reader, and refer to the original works for details of the
derivation~\cite{gardiner93a,carmichael93b} or to the Supplementary
Material of Ref.~\onlinecite{lopezcarreno15a}, where it is cast in the
problematic of the present
text. Section~\ref{sec:lunnov30100948CET2015} gives an overview of the
many possibilities one can study as well as details of the
configurations we focus on in the following of the series. After what
can be seen as a long introduction,
Section~\ref{sec:lunnov30101947CET2015} introduces the first important
results by providing a way to characterize quantum states of the
Harmonic oscillator in a space of quantum-optical diagonal
correlators, which will be helpful to later characterize the sources,
and with effect to disprove a popular criterion for single-photon
states.  Section~\ref{sec:vieoct23172746CEST2015} and
\ref{sec:marnov24163921CET2015} characterize the incoherent and
coherent SPS, respectively. As the latter will prove to be more
interesting, we devote most of our attention to this case.
Section~\ref{sec:lunnov30102926CET2015} shows that the advantages of
the cascaded coupling over the conventional Hamiltonian coupling
(cavity QED) remain present even when considering alternative
descriptions of coupling between the source and the target.  Section
\ref{sec:jueoct29112554CET2015} draws the conclusions for the cases
studied here while Section~\ref{sec:jueoct29112601CET2015} does so for
the wider picture of cascaded coupling and introduces the other cases
to be investigated in follow-up papers.

\begin{figure}
    \includegraphics[width=0.8\linewidth]{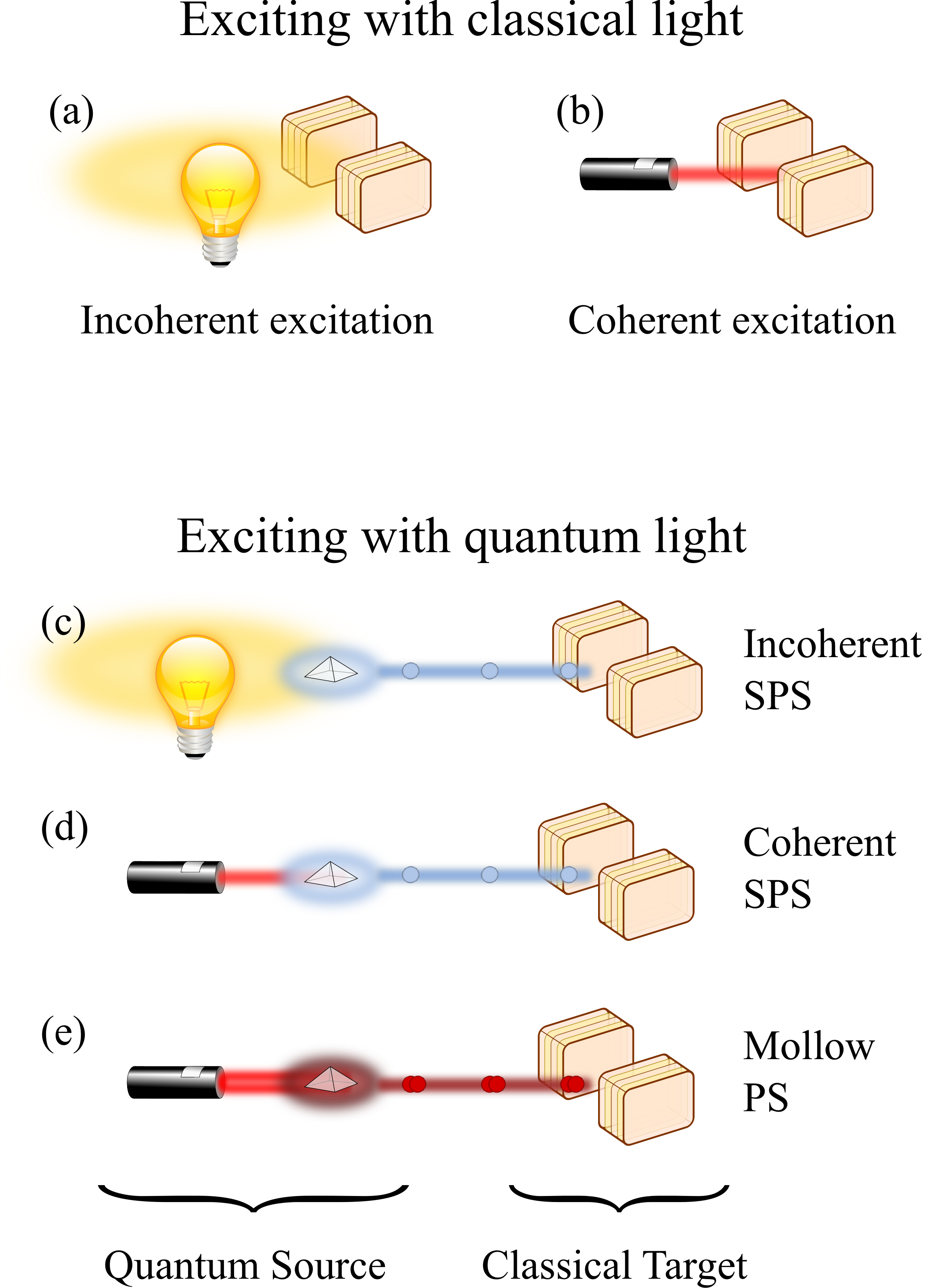}
    \caption{(Color online). Upper part: The two main types of
      classical excitation. The harmonic oscillator can be implemented
      by a single-mode cavity, sketched here as two distributed Bragg
      reflectors facing each other. The SPS can be implemented by a
      quantum dot, sketched here as a little pyramid, as it would be
      grown by self-assembly~\cite{grundmann95a}. (a) Incoherent
      excitation, typically corresponding to thermal light. (b)
      Coherent excitation, typically corresponding to driving the
      system with a laser. Lower part: Upgrading the classical sources
      of the upper panel with SPS. (c) is the counterpart of~(a), to
      be referred to as the ``incoherent SPS''. The coherent case
      yields two very different types of quantum light at low and high
      pumpings: (d) the coherent SPS and (e) the Mollow PS which can
      emit more than one photon at a time.  Other types of quantum
      sources and other types of targets constitute the topic of the
      following texts in the series.}
  \label{fig:jueoct29162857CET2015}
\end{figure}

\section{Theoretical background}
\label{sec:marnov24124242CET2015}

The coupling between two quantum systems is typically given by an
interaction Hamiltonian. In the second quantization formalism, such a
Hamiltonian reads in its most simple form (we take $\hbar =1$ along
the paper):
\begin{equation}
  H_{I} = g(\ud{c_1}c_2+\ud{c_2}c_1)\,,
  \label{Eq.InteractionHamiltonian}
\end{equation}
where $c_1$, $c_2$ are annihilation operators describing the particles
of the coupled system, and $g$ is their interaction strength. If the
particles described by $c_i$ have a decay rate $\gamma_i$, and are
freely evolving with a Hamiltonian $H_i$, $i=1,2$, the master equation
describing the dynamics of this system is:
\begin{equation}
\partial_t \rho = i [\rho, H_1 +H_2+H_I] + \frac{\gamma_1}{2}\mathcal{L}_{c_1}\rho+\frac{\gamma_2}{2}\mathcal{L}_{c_2} \rho\, ,
\label{Eq.MasterEquationNormal}
\end{equation}
where
$\mathcal{L}_c \rho = (2 c\rho \ud{c} - \ud{c}c\rho - \rho \ud{c}c)$.
Depending on whether the coupling~$H_I$ or the dissipation~$\gamma_i$
dominates, one speaks of strong or weak coupling, respectively. If one
of the two systems, say, 1, is itself excited externally, for instance
being driven by a laser, or merely being given a non-vacuum initial
condition, then one has a crude picture of system~1 exciting
system~2. This is a fairly accurate description in the weak-coupling
limit where the dynamics seems irreversible, simply because
excitations are dissipated before they can cycle back (we discuss in
Section~\ref{sec:lunnov30102926CET2015} when this becomes exact).

The coupling between quantum systems does not have to be reversible:
it can instead correspond to the scenario of a source and its target.
In this case, there is a deep asymmetry between the coupled
systems. For instance, one can remove the target from the beam of the
source, which leaves the latter unaffected while the former passes
from being irradiated to the vacuum. Note that such an asymmetry does
not have to hold on logical grounds. In fact, in electronics, while an
ideal source should not be affected by the circuit it powers, in
reality, there is a load and every component affects all the others to
some extent. In photonics, the picture of a flying qubit, left to
propagate long enough before it meets its target, makes it intuitively
clear that it should be possible to forbid back-action. This could
also be realized by taking advantage of the fast-growing field of
chiral optics.\cite{petersen14a, gonzalezballestero15a, pichler15a,
  coles16a, arXiv_sorensen16a, arXiv_guimond16a, arXiv_lodahl16a,
  arXiv_gonzalezballestero16a} Theoretically, this asymmetry is
achieved by the cascaded coupling.\cite{gardiner_book00a}, where the
equations of motion are expressed in the quantum Langevin form, thus
allowing to set the output field of one of the systems (the source) as
the input field for the other (the target). This can be brought to a
master equation type of description, with both coherent and Lindblad
terms that contrive to direct the flow of excitation from the source
to target only. This makes all the operators of the source independent
from those of the target, while in turn those depend on operators of
the source. The generic case where the source (resp.~target) is
described by the Hamiltonian $H_1$ (resp.~$H_2$) and has a decay
rate~$\gamma_1$ (resp.~$\gamma_2$) is ruled by the following master
equation:\cite{gardiner_book00a}
\begin{multline}
  \partial_t\rho = i [\rho, H_{1}+H_{2}] +
  \frac{\gamma_1}{2}\mathcal{L}_{c_1} \rho +
  \frac{\gamma_2}{2}\mathcal{L}_{c_2} \rho-{} \\{}- \sqrt{\gamma_1
    \gamma_2}\left ( [\ud{c_2},c_1\rho]+[\rho\ud{c_1}, c_2] \right)\, .
\label{Eq.MasterEquationCascaded}
\end{multline}
The source must also be excited, which can be done either by an
incoherent or by a coherent (classical) type of excitation. The
incoherent excitation is described simply by adding the Lindblad term
$(P_{c_1}/2) \mathcal{L}_{\ud{c_1}}\rho$ to
Eq.~(\ref{Eq.MasterEquationCascaded}). The coherent excitation,
however, requires a subtler description, for which one uses the
input--output formalism. The coupling between a coherent field and the
system (that latter will be used as the source of quantum excitation)
happens through an input channel for the said system. If such an input
channel is the only one available to excite the source, then it
follows that the only output channel from the source also contains the
coherence of the driving field. In this case, the target of the
quantum exitation is \emph{also} driven by the coherent field, and its
dynamics is given by Eq.~(\ref{Eq.MasterEquationCascaded}) setting
$H_1=\omega_1
\ud{c_1}c_1-i\sqrt{\gamma_{1}}\mathcal{E}(\ud{c_1}-c_1)$,
and
$H_2=\omega_2
\ud{c_2}c_2-i\sqrt{\gamma_{2}}\mathcal{E}(\ud{c_2}-c_2)$,
i.e., the dynamics is ruled by the master equation:
\begin{multline}
  \label{Eq.MasterEquationCoherent1}
  \partial_t\rho = i \left[\rho, \omega_1\ud{c_1}c_1+\omega_2\ud{c_2}c_2-i\sqrt{\gamma_{1}}\mathcal{E}(\ud{c_1}-c_1)\right.-{}\\
{}-\left. i\sqrt{\gamma_{2}}\mathcal{E}(\ud{c_2}-c_2)\right]+
  \frac{\gamma_1}{2}\mathcal{L}_{c_1} \rho +
  \frac{\gamma_2}{2}\mathcal{L}_{c_2} \rho-{} \\{}- \sqrt{\gamma_1
    \gamma_2}\left ( [\ud{c_2},c_1\rho]+[\rho\ud{c_1}, c_2] \right)\,,
\end{multline}
where $\mathcal{E}$ is the amplitude of the coherent field driving the
source. Note that the effective driving intensity, i.e.,
$\Omega_1=\sqrt{\gamma_{1}}\mathcal{E}$, depends on the decay rate of
the system that is being excited, in agreement with the fact that a
system that cannot emit cannot be excited either. To prevent the
target to be also driven by the coherent field, one can use other
input (and their corresponding output) channels to excite the source
(and also the target). Each of these channels couples with an
amplitude $\epsilon_i\leq 1$, with the condition that
$\sum_{k}\epsilon_k=1$, the sum being over all the input channels. In
this case, and considering only two input channels (with amplitudes
$\epsilon_1$ and $\epsilon_2=1-\epsilon_1$) as well as only one input
channel for the target (with amplitude $1$), the dynamics of the
system is given by Eq.~(\ref{Eq.MasterEquationCascaded}) with
$H_1=\omega_1
\ud{c_1}c_1-i\sqrt{\epsilon_1\gamma_1}\mathcal{E}(\ud{c_1}-c_1)$,
$H_2=\omega_2 \ud{c_2}c_2$, and replacing the coupling strength
$\sqrt{\gamma_1 \gamma_2}$ by $\sqrt{(1-\epsilon_1)\gamma_1 \gamma_2}$
in the second line of Eq.~(\ref{Eq.MasterEquationCascaded}), i.e., the
dynamics is now ruled by the master equation:
\begin{multline}
  \label{Eq.MasterEquationCoherent2}
  \partial_t\rho = i \left[\rho, \omega_1\ud{c_1}c_1 +\omega_2\ud{c_2}c_2-i\sqrt{\epsilon_1 \gamma_{1}}\mathcal{E}(\ud{c_1}-c_1)\right]+{}\\
{} +\frac{\gamma_1}{2}\mathcal{L}_{c_1} \rho +
  \frac{\gamma_2}{2}\mathcal{L}_{c_2} \rho- \sqrt{\epsilon_2\gamma_1
    \gamma_2}\left ( [\ud{c_2},c_1\rho]+[\rho\ud{c_1}, c_2] \right)\,,
\end{multline}
where $\epsilon_2=1-\epsilon_1$. The additional input channel to the
source makes the coupling between the coherent field not as efficient
as when there is only one input channel, thus reducing the effective
driving intensity. For the target, although the coupling strength is
also reduced, now the driving is uniquely due to the emission from the
quantum source.

Putting Eq.~(\ref{Eq.MasterEquationCascaded}) in the Lindblad form
contributes a Hamiltonian part.  The formalism thus corresponds to a
quantum coherent coupling, allowing the description of continuous wave
(cw) and resonant excitation of quantum states.  Importantly, in
contrast to the Hamiltonian coupling in
Eq.~(\ref{Eq.InteractionHamiltonian}), the coupling strength is now
fixed by the decay rate of the source and of the target. An
infinitely-lived target cannot be excited. The stronger one wishes to
make the coupling between a source and its target, the stronger has to
be their (geometric) mean dissipation. This imposes some fundamental
constrains on external driving (or driving without feedback). In
contrast, Hamiltonian coupling sets the coupling strength and decay
rates independently. While it would therefore appear that the
Hamiltonian coupling has the upper hand, and that one should strive
for the standard strong-coupling regime, we will show in the following
that the cascaded architecture can be superior to the other types of
coupling in some cases.

\section{Quantum Sources and Optical Targets}
\label{sec:lunnov30100948CET2015}

The general problem of excitation with quantum light has obviously
numerous ramifications. To the already large variety of optical
targets, one now has to combine a much enlarged set of quantum
sources. This literally opens a new dimension to optics.  Indeed,
classical excitation could arguably be limited to a rather small set
of categories:
\begin{enumerate}
\item[(a)] Coherent excitation (driving with a laser)
\item[(b)] Incoherent excitation (incoherent pumping, Boltzmann
  dynamics, thermal baths, equilibrium, etc.)
\end{enumerate}
(we postpone to part~V the discussion of time-dependent and pulsed
excitation and consider until then the case of continuous wave
excitation, cf.~Section~\ref{sec:jueoct29112601CET2015}). Quantum
light, on the other hand, encompasses not only the above---if only
because it is a more general case that includes classical excitation
as a particular case---but also comes with many more and higher
degrees of freedom. This did not lead so far, to the best of our
knowledge, to a classification. Tentatively, this could be provided in
a first approximation by~$g^{(2)}$ (antibunching, uncorrelated,
bunching, superbunching). Since squeezed states have the correlations
of coherent states, however, and they are precisely the type of input
that motivated a new formalism, it is clear that this is still far
from appropriate. We decided to approach this general problem by
considering:
\begin{itemize}
\addtolength{\itemsep}{-0.5\baselineskip}
\item the same optical target (here an harmonic oscillator, in the next
  paper a two-level system),
\item the same type of quantum source (here and in the next paper, a SPS, in the third paper, an $N$-photon emitter),
\item both coherent and incoherent regimes,
\item both low and high pumping regimes.
\end{itemize}

\begin{table*}[thpb]
  \begin{ruledtabular}
    \begin{tabular}{l|c|c|c|c|c}
      Source & Pumping & Population~$n_\sigma$ & $g_\sigma^{(2)}(\tau)$ & Lineshape & Linewidth \\
      \hline
      Incoherent SPS & $P_\sigma$ & $P_\sigma/(\gamma_\sigma+P_\sigma)$ & $1-\exp\big(-(\gamma_\sigma+P_\sigma)\tau\big)$ & Lorentzian & $\gamma_\sigma+P_\sigma$ \\
      Coherent SPS & small $\Omega_\sigma$ & $4\Omega_\sigma^2/\gamma_\sigma^2$ & $(1-\exp(-\gamma_\sigma\tau/2))^2$ & Lorentzian & $\gamma_\sigma$ \\
      Coherent SPS & large $\Omega_\sigma$ & $4\Omega_\sigma^2/(\gamma_\sigma^2+8\Omega_\sigma^2)$ & $\displaystyle 1-e^{-\frac{3\gamma_\sigma\tau}{4}}\Big[\cosh(\frac{\Gamma_\sigma\tau}{4})+\frac{3\gamma_\sigma}{\Gamma_\sigma}\sinh(\frac{\Gamma_\sigma\tau}{4})\Big]$ & Triplet & $\left\{\pbox{20cm}{$\gamma_\sigma$ (central)\\$3\gamma_\sigma/2$ (satellites)}\right.$ \\
    \end{tabular}
  \end{ruledtabular}
  \caption{Characteristic of a SPS driven incoherently or coherently. In the latter case, for clarity, we also consider separately the low-driving intensity (middle row) from the general case (bottom row) that covers all pumpings, to highlight the strong qualitative features of the Mollow regime. \label{tab:vieoct23142648CEST2015}}
\end{table*}

The SPS is a good starting point because it is the paradigm of quantum
light and is the most common light of this type in the
laboratory. Exciting the harmonic oscillator makes the problem both
simple and fundamental, so this is also a good starting point. While
we have considerably restrained the possibilities, there still remains
much to be explored and the following will only address the main
results. The two types of excitations aforementionned~(a--b) yield
quite different types of SPS, that are introduced in the beginning of
their respective Sections, and are summarized in
table~\ref{tab:vieoct23142648CEST2015}. Note that in the table, the
last row also includes the second one (which is the limit of small
pumping). They are however so distinct qualitatively that it is
helpful to think of them as separate sources. As we will show, the
second row (small coherent pumping) leads to the best antibunching in
the target.

\section{Charting the Hilbert space}
\label{sec:lunnov30101947CET2015}

Prior to considering which quantum states of the harmonic oscillator
one can access by exciting it with various types of SPS, one needs a
roadmap to characterize all of the states at a glance. The Hilbert
space is a big place. Human's capacity of abstraction gives a
deceivingly simple picture of it. The canonical basis of Fock
states~$\ket{n}$ with~$n\in\mathbb{N}$ provides a comprehensive and
concise map of the states in the harmonic oscillator's Hilbert
space~$\mathcal{H}_\infty$:
\begin{equation}
  \label{eq:miéoct21094346CEST2015}
  \ket{\psi}=\sum_{n=0}^\infty\alpha_n\ket{n}\,,
\end{equation}
with~$\alpha_n\in\mathbb{C}$ such that
$\sum_{n=0}^\infty|\alpha_n|^2=1$. This is as precise as it is
misleading, since this fails to provide a qualitative map of the possible
states. Instead, one uses families of particular cases. Beyond Fock
states per se, one makes a great use of the coherent states of
classical physics, for
which~$\alpha_n=\alpha^n\exp(-|\alpha|^2)/\sqrt{n!}$ for some complex
number~$\alpha$ that defines the amplitude and phase of a classical
field. The next family of important states requires the concept of
mixity that involves statistical averages and upgrades the
wavefunction to a density matrix:
\begin{equation}
  \label{eq:miéoct21095538CEST2015}
  \rho=\sum_{n,m=0}^\infty\beta_{nm}\ket{n}\bra{m}\,.
\end{equation}
%
This allows the introduction of thermal states, for which
$\beta_{nm}=(1-\theta)\theta^n\delta_{nm}$ for some reduced
temperature~$\theta\in[0,1]$. These two families can be united into a
larger class of ``cothermal''\cite{laussy06a} states that interpolate
between the two and describe, e.g., single-mode Bose condensates not
too far from threshold. From there on, one essentially deals with
quantum states of light with popular examples such as squeezed
states,\cite{lvovsky_inbook15a} cat states,\cite{haroche_book06a}
etc., and other less popular such as binomial states\cite{stoler85a}
or displaced Fock states.\cite{nieto97a} Such a zoology of states is
familiar to every quantum physicist, but it fails to provide the
sought mapping of the Hilbert space. In particular, it is restricted
to known (and popular) cases and leaves most of
Eqs.~(\ref{eq:miéoct21094346CEST2015}--\ref{eq:miéoct21095538CEST2015})
uncharted.

We introduce a map of the Hilbert space based on important observables
for the quantum state, namely, the $n$-th order correlation functions:
\begin{equation}
\label{eq:jueoct29094202CET2015}
g^{(n)}=\frac{\langle a^{\dagger n}a^n\rangle}{\langle\ud{a}a\rangle^n}\,.
\end{equation}
Since we consider only the quantum states and not their dynamics,
normalization makes $g^{(1)}$ trivially unity. We therefore use
instead the normalization itself, which is an important
observable, the population of the state (average number of
excitations), for which we introduce the notation:
\begin{equation}
  \label{eq:jueoct29094507CET2015}
  n_a\equiv\langle\ud{a}a\rangle\,.
\end{equation}
Note that only the diagonal elements~$p(m)=|\alpha_m|^2$ are needed so
there is no distinction between pure and mixed states in this
discussion. The $n$-th order correlator reads:
\begin{equation}
  \label{eq:jueoct29095412CET2015}
  g^{(n)}=\frac{\sum_{m=0}^\infty m!p(m)/(m-n)!}{\big(\sum_{m=0}^\infty mp(m)\big)^n}\,.
\end{equation}

\begin{figure}
  \includegraphics[width=0.95\linewidth]{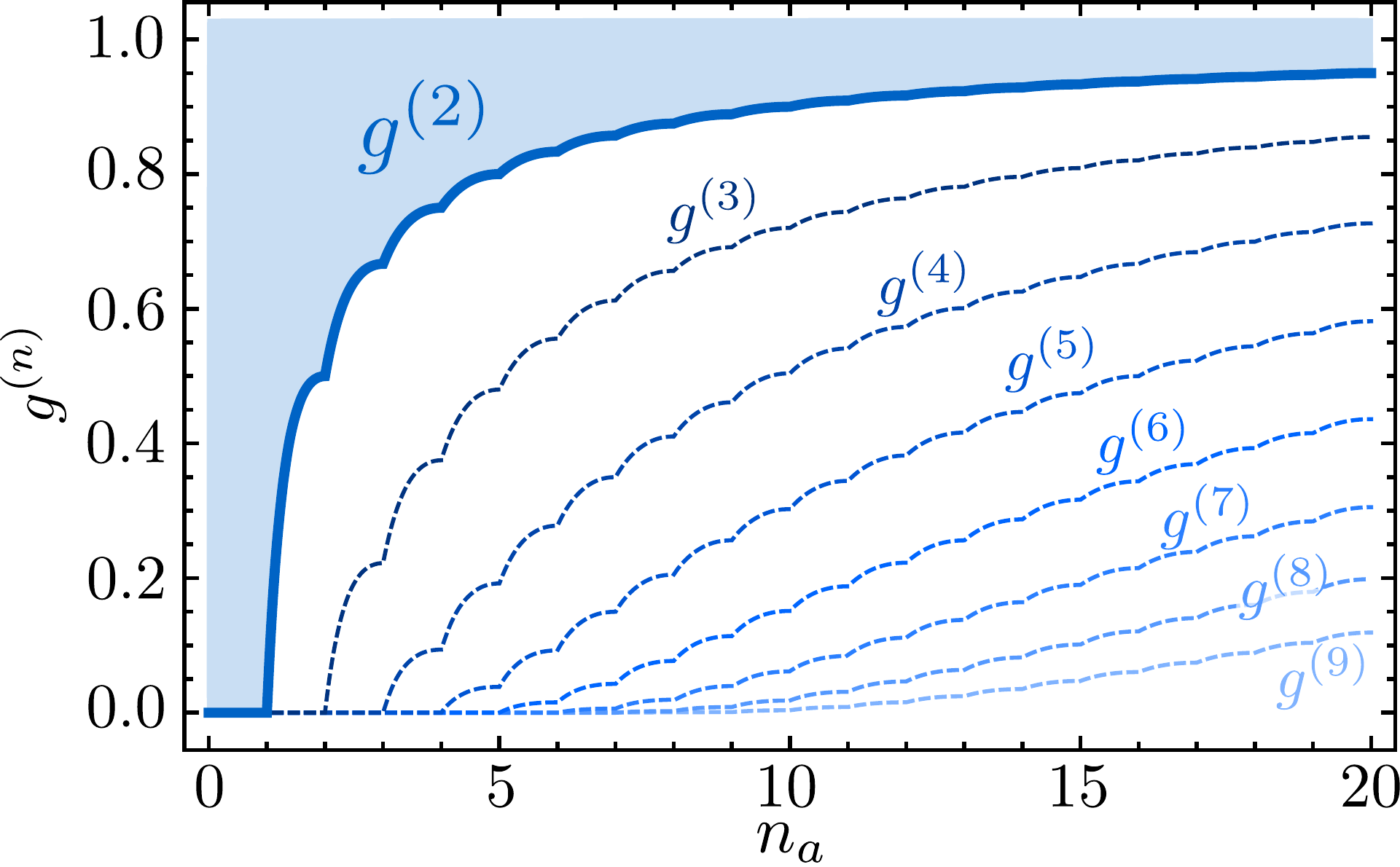}
  \caption{(Color online). Charting the Hilbert space. The shaded area
    shows the region where a physical quantum state exists with the
    corresponding joint population and antibunching. It is delimited
    by the boundary Eq.~(\ref{eq:miéoct21181551CEST2015}) of ``Fock
    duos'', defined by Eq.~(\ref{eq:marnov24092156CET2015}).  Higher
    order correlators (thin dashed) set other boundaries, also
    delimited by Fock duos.}
  \label{fig:miénov25113249CET2015}
\end{figure}

A convenient mapping of the Hilbert space is to chart it with the
$n_a$ and $g^{(2)}$ ``rulers'', that is, tag the possible quantum
states through their joint statistical properties and population. The
value of~$g^{(2)}$ allows to tell apart classical from quantum states
depending on whether $g^{(2)}$ is larger (bunching) or smaller
(antibunching) than unity. This is meant in the sense of whether a
classical, possibly stochastic, description is possible, or whether
some ``quantum'' features such as violation of Cauchy--Schwarz
inequalities or negative probabilities in the phase space are
manifest. The population gives another meaning to quantumness, in the
sense of few-particle effects vs macroscopic occupation. It is
therefore particularly interesting to contrast these two attributes
and consider, e.g., highly occupied genuinely quantum states.  There
is a physical limit to such states and for populations~$n_a>1$, some
values of antibunching are out of reach.  This is expected on physical
grounds from familiar features of the Fock states popularly known as
being the ``most quantum'' states, with their $g^{(2)}$ tending to one
as the number of excitations increases. Also, macroscopic quantum
states, such as a BEC, are essentially coherent states in most
formulations. Here we make this notion precise by providing the
complete picture, in Fig.~\ref{fig:miénov25113249CET2015}, along with
the closed-form expression for the boundary that separates accessible
combinations from impossible ones (this is proved in
Appendix~\ref{sec:marnov24122827CET2015}):
\begin{equation}
  \label{eq:miéoct21181551CEST2015}
  g^{(2)}=\frac{\lfloor n_a\rfloor(2n_a-\lfloor n_a\rfloor-1)}{n_a^2}\,.
\end{equation}
This boundary is provided by superpositions of contiguous Fock states,
i.e., of the type:
\begin{equation}
  \label{eq:marnov24092156CET2015}
  \sqrt{p(n)}\ket{n}+\sqrt{1-p(n)}e^{i\theta}\ket{n+1}\,,
\end{equation}
for~$n\in\mathbb{N}$, $p(n)\in[0,1]$ and~$\theta\in[0,2\pi[$ an
(irrelevant) phase. (As already commented, this includes also mixed
states of the type $p(n)\ket{n}\bra{n}+(1-p(n))\ket{n+1}\bra{n+1}$ and
all others with the same diagonal but different off-diagonal
elements. We will not make this distinction anymore in the following.)
We call these states, Eq.~(\ref{eq:marnov24092156CET2015}),
``\emph{Fock duos}''.  They set the continuous lower limit of the
space of available quantum states in our charted Hilbert space. Their
antibunching~(\ref{eq:miéoct21181551CEST2015}) is a generalization to
non-integer~$n_a$ of the formula $g^{(2)}=1-1/n_a
$ for the Fock state~$\ket{n_a}$ to which it reduces
for~$n_a\in\mathbb{N}$.  It shows that the popular ``single-particle
criterion'' that asserts that $g^{(2)}<0.5$ ensures a one particle
state~\cite{michler00a,dong07a,verma11a,dimartino12a,leifgen14a} is
wrong, as demonstrated, e.g., for the case~$n_a=1.5$ for which
$g^{(2)}=4/9\approx0.44$.  The generalization to higher orders is
straigthforward. Namely, the $n$th-order correlation frontier for
real~$n_a$ is given by:
\begin{equation}
  \label{eq:jueoct29102628CET2015}
  g^{(n)} =\frac{\lfloor n_a\rfloor!}{(\lfloor n_a\rfloor-n)!n_a^n}\left [1+\frac{n(n_a-\lfloor n_a\rfloor)}{\lfloor n_a\rfloor+1-n} \right ]\,,
\end{equation}
that generalizes the Fock state formula $g^{(n)}=(1/n_a^n)m!/(m-n_a)!$
valid for integer~$n_a$ where it agrees with
Eq.~(\ref{eq:jueoct29102628CET2015}). Interestingly, the lower limit
is also set, for all~$n$, by Fock duos (not ``$n$-plets'').

One can also consider other charts of the Hilbert space, such as
$(g^{(2)}, g^{(3)})$, this time contrasting two- and three-particle
correlations together. This time there are no boundaries in this space
(one can find state with any joint values of $g^{(2)}$ and
$g^{(3)}$). The proof of this statement is given in
Ref.~\onlinecite{arxiv_zubizarreta16a}. Since the excitation of an
harmonic oscillator by the SPS leads to strong correlations between
various $g^{(n)}$, we will focus on the $(n_a,g^{(2)})$ space.

\section{Exciting with an Incoherent SPS}
\label{sec:vieoct23172746CEST2015}

Now that we can conveniently characterize the states of the harmonic
oscillator that one can excite, we come back to the dynamical problem
of the quantum driving of an harmonic oscillator and how close can one
get to the limit set by Eq.~(\ref{eq:miéoct21181551CEST2015}).  We
first consider the case of excitation with an incoherent SPS, i.e.,
where the two-level system (2LS) that acts as the source, is itself
driven incoherently, as sketched in
Fig.~\ref{fig:jueoct29162857CET2015}(c). Namely, there is a constant
rate~$P_\sigma$ at which the 2LS is put in its excited state, and is
otherwise left to decay. The system is thus described by the master
equations~(\ref{Eq.MasterEquationNormal}--\ref{Eq.MasterEquationCascaded}),
with $c_1=\sigma$ the 2LS and $c_2=a$ the harmonic oscillator
 operators, respectively. The 2LS pumping is described by
a Lindblad term $(P_\sigma/2)\mathcal{L}_{\ud{\sigma}}\rho$. Thanks to
the absence of feedback, the dynamics is ruled by closed equations
which allows us to obtain exact solutions for the observables of
interest. Namely, we find that the cavity population~$n_a$ and
statistics~$g^{(2)}$ of the target are given by
(cf.~Table~\ref{tab:vieoct23142648CEST2015} for the source):
\begin{subequations}
  \label{eq:juedic17123703CET2015}
  \begin{align}
    n_a &= \frac{4 P_\sigma \gamma_\sigma}{(\gamma_\sigma+P_\sigma)(
        \gamma_\sigma+P_\sigma+\gamma_a)}\,,\label{eq:vieene15123031CET2016}\\
    g^{(2)} &=
    \frac{2(\gamma_\sigma+P_\sigma)}{\gamma_\sigma+P_\sigma+3\gamma_a}\,.\label{eq:vieene15123037CET2016}
  \end{align}
\end{subequations}

\begin{figure}
    \includegraphics[width=0.95\linewidth]{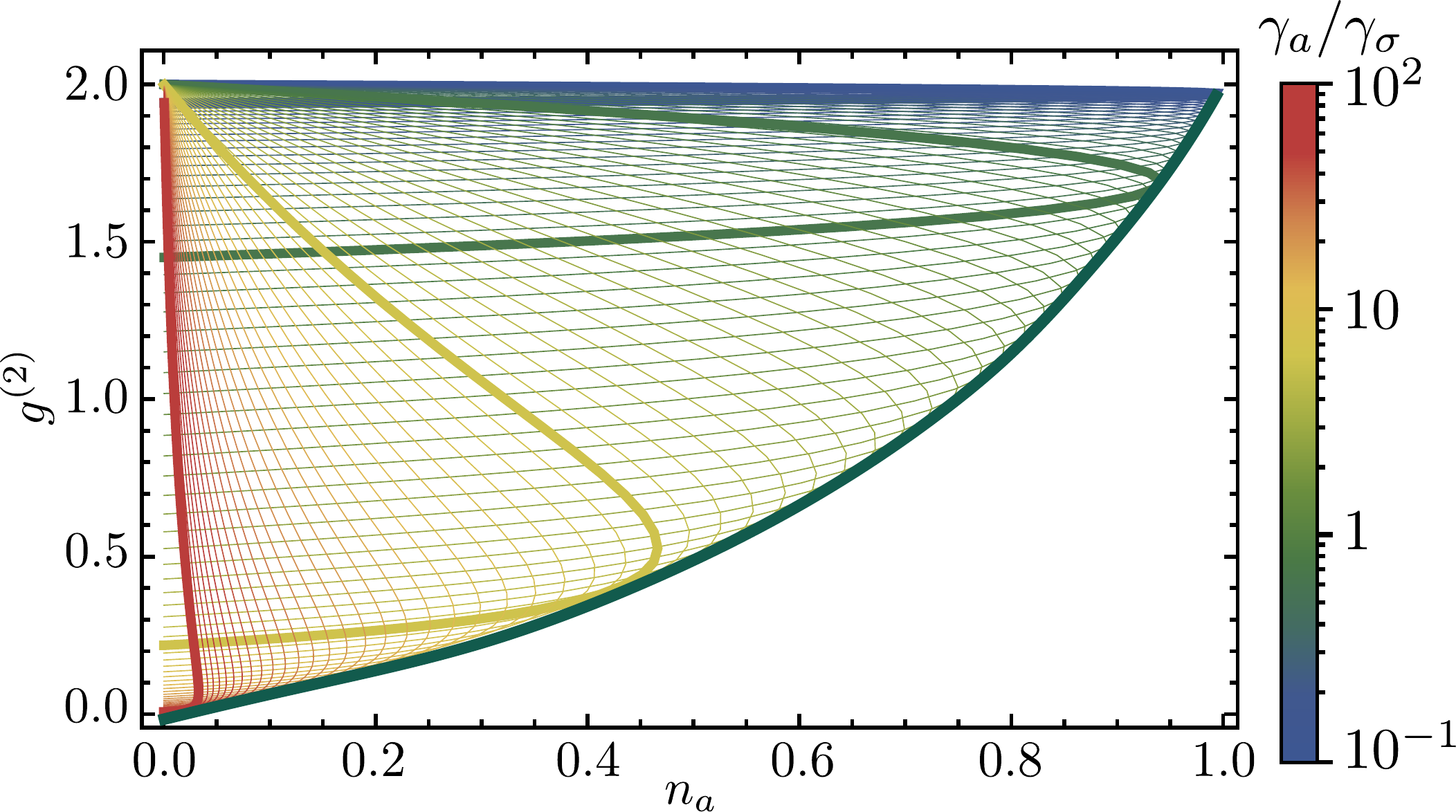}
    \caption{(Color online). Trajectories in the Hilbert space charted
      by~$n_a$ and~$g^{(2)}$ for the states excited by an incoherent
      SPS, for various values of~$\gamma_a/\gamma_\sigma$. The
      implicit parameter is pumping. Highlighted are the
      cases~$\gamma_a/\gamma_\sigma=10^{-1}$ (blue), 1 (green), 10
      (yellow) and $10^2$ (red). The dark green thick envelope shows
      the closest one can get in this configuration to the ideal
      antibunching, which is zero. All the states, and only these
      states, above this line and below~$g^{(2)}=2$, are accessible
      with an incoherent SPS.}
  \label{fig:juedic17170852CET2015}
\end{figure}

Only two parameters are required to describe fully the system: i)~the
ratio between the decay rate of the cavity and the emission rate of
the 2LS, $\gamma_a/\gamma_\sigma$, and ii)~the intensity (or pumping
rate) of the 2LS, $P_\sigma$, also normalized to $\gamma_\sigma$ to
keep the parameters dimensionless. Eliminating~$P_\sigma$ from
Eqs.~(\ref{eq:juedic17123703CET2015}), this gives the equation for the
trajectory in the $(n_a,g^{(2)})$ space as function of the
parameter~$\gamma_a/\gamma_\sigma$:
\begin{equation}
  \label{eq:juedic17155545CET2015}
  n_a=\frac23\frac{(2-g^{(2)})(3g^{(2)}(\gamma_a/\gamma_\sigma)-(2-g^{(2)}))}{g^{(2)}(1+g^{(2)})(\gamma_a/\gamma_\sigma)}\,.
\end{equation}
These trajectories in the Hilbert space are plotted in
Fig.~\ref{fig:juedic17170852CET2015}. The curves start from the point:
\begin{equation}
  \label{eq:juedic17173145CET2015}
  \left(n_a=0,g^{(2)}=\frac{2}{1+3\gamma_a/\gamma_\sigma}\right)\,,
\end{equation}
at vanishing pumping, reach a turning point with coordinates:
\begin{multline}
  \label{eq:juedic17174537CET2015}
  \left(n_a=\frac{4(2+(\gamma_a/\gamma_\sigma)-2\sqrt{1+(\gamma_a/\gamma_\sigma)})}{(\gamma_a/\gamma_\sigma)^2},\right.\\
  \left.g^{(2)}=\frac{2}{3\sqrt{1+\gamma_a/\gamma_\sigma}-2}\right)\,,
\end{multline}
when $P_\sigma=\sqrt{\gamma_\sigma (\gamma_a + \gamma_\sigma)}$ and
converge at~$(n_a=0,g^{(2)}=2)$ at large pumpings, where the source
gets quenched. For each value of~$\gamma_a/\gamma_\sigma$, there are
two values of pumping that result in the same population but two
values of~$g^{(2)}$.  The curve of
Eq.~(\ref{eq:juedic17155545CET2015}) is fairly constant till the
turning point and, from Eq.~(\ref{eq:juedic17173145CET2015}), leads to
a genuine quantum state (i.e., featuring antibunching) as long
as~$\gamma_a/\gamma_\sigma>1/3$, which means that, with a SPS, one can
imprint antibunching in a system that has a substantially longer
lifetime.  The optimum antibunching/population for a given
$\gamma_a/\gamma_\sigma$ is achieved slightly earlier than the turning
point, namely, when $P_\sigma = \gamma_\sigma$. The envelope of all
the curves in Fig.~\ref{fig:juedic17170852CET2015} is the closest one
can get to the Fock-duos limit, and is given by
$g^{(2)} = 2 n_a /(3-2n_a)$. This is the thick dark green line in
Fig.~\ref{fig:juedic17170852CET2015}.

\begin{figure*}
    \includegraphics[width=0.95\linewidth]{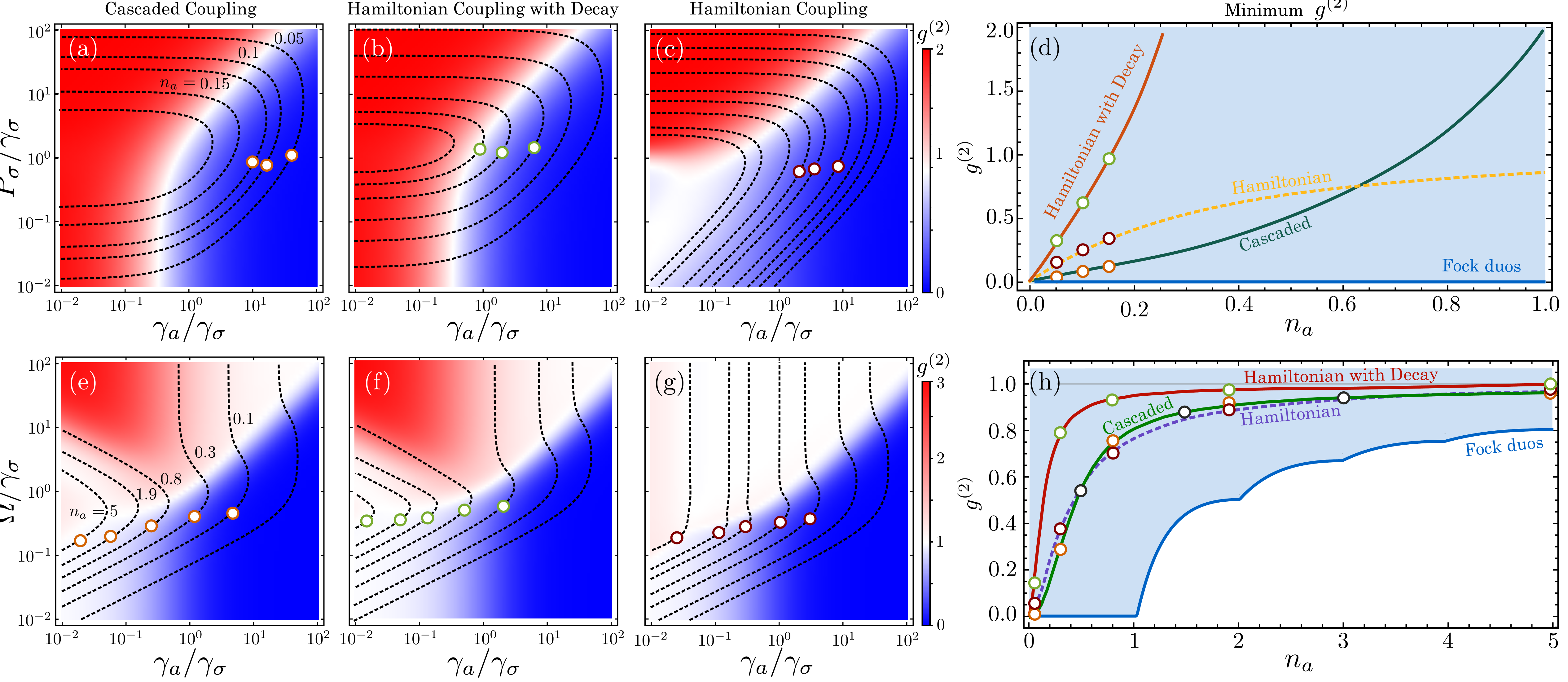}
    \caption{(Color online). Upper row: Populations (isolines) and
      $g^{(2)}$ (color) of the target excited by an incoherent SPS,
      through (a) the cascaded coupling or (b,c) the Hamiltonian
      coupling with (b) and without (c) decay of the 2LS. (d) Minimum
      $g^{(2)}$ that can be reached at a given population for all
      these cases.  The color circles mark the parameters
      $P_\sigma/\gamma_\sigma$ and $\gamma_a/\gamma_\sigma$ at which
      the population isoline meets the minimum $g^{(2)}$.  Lower row:
      Same but for a coherent SPS, with (e) the cascaded coupling and
      (f,g) the Hamiltonian coupling with (f) and without (g) decay of
      the 2LS.  (h) Minimum $g^{(2)}$ that can be reached at a given
      population for all these cases. The color circles mark the
      parameters $\Omega_\sigma/\gamma_\sigma$ and $\gamma_a
      /\gamma_\sigma$ at which the population isoline meets the
      minimum $g^{(2)}$.}
  \label{fig1}
\end{figure*}

The above solution gives an already substantial description of the
response of an harmonic oscillator to an incoherent SPS. We can
complement it with alternative descriptions that approach the solution
from different viewpoints and manifest the advantage of cascaded
coupling over other types of excitation. A natural comparison is with
the standard Hamiltonian coupling, that is the usual way to couple
different systems.  For instance, it is convenient to lay out the
observables as function of the parameters in 2D~density plots, as
shown in the upper row of Fig.~\ref{fig1}. Panels~(a--c) show the
photon statistics (color code) and the population of the cavity
(isolines) using the (a) cascaded coupling and (b, c) the Hamiltonian
coupling. For the later case, we consider two configurations: (b) when
the 2LS has the same decay term as when acting as a source in the
cascaded scheme, and (c) when the 2LS has no decay term, so that all
its excitation is directed towards the cavity, corresponding to the
ideal excitation of the cascaded scheme. Interestingly, (a) and~(b)
types of excitation present a qualitatively similar photon statistics
layout, but the topography of the resulting population yields largely
differing states, as evident once reported in the $(n_a,g^{(2)})$
space, panel~(d).  The color circles along the isolines mark the
points with strongest antibunching (i.e., smallest~$g^{(2)}$) for a
given population. It is seen that for a given signal, the antibunching
is significantly larger with the cascaded coupling. In both cases the
antibunching is marred before the cavity population reaches one photon
on average. At larger populations, the photon statistics of the cavity
become bunched, i.e., $g^{(2)}=2$.  On can contemplate other
configurations but they result in worse results: detuning the cavity
from the 2LS leaves the photon statistics unchanged, but reduces the
population of the cavity, so the antibunching in lost at even smaller
cavity populations. Dephasing the 2LS increases the broadening of the
emission peak, but also drives the cavity towards a coherent state
faster.  Hamiltonian coupling with no decay of the 2LS, case~(c),
differs qualitatively from the former cases and provides a
substantially better antibunching than Hamiltonian coupling with decay
of the 2LS. It even becomes better than the cascaded coupling when
$n_a\gtrsim 2/3$.  This situation can be much improved with the
coherent SPS.

\begin{figure*}
    \includegraphics[width=0.95\linewidth]{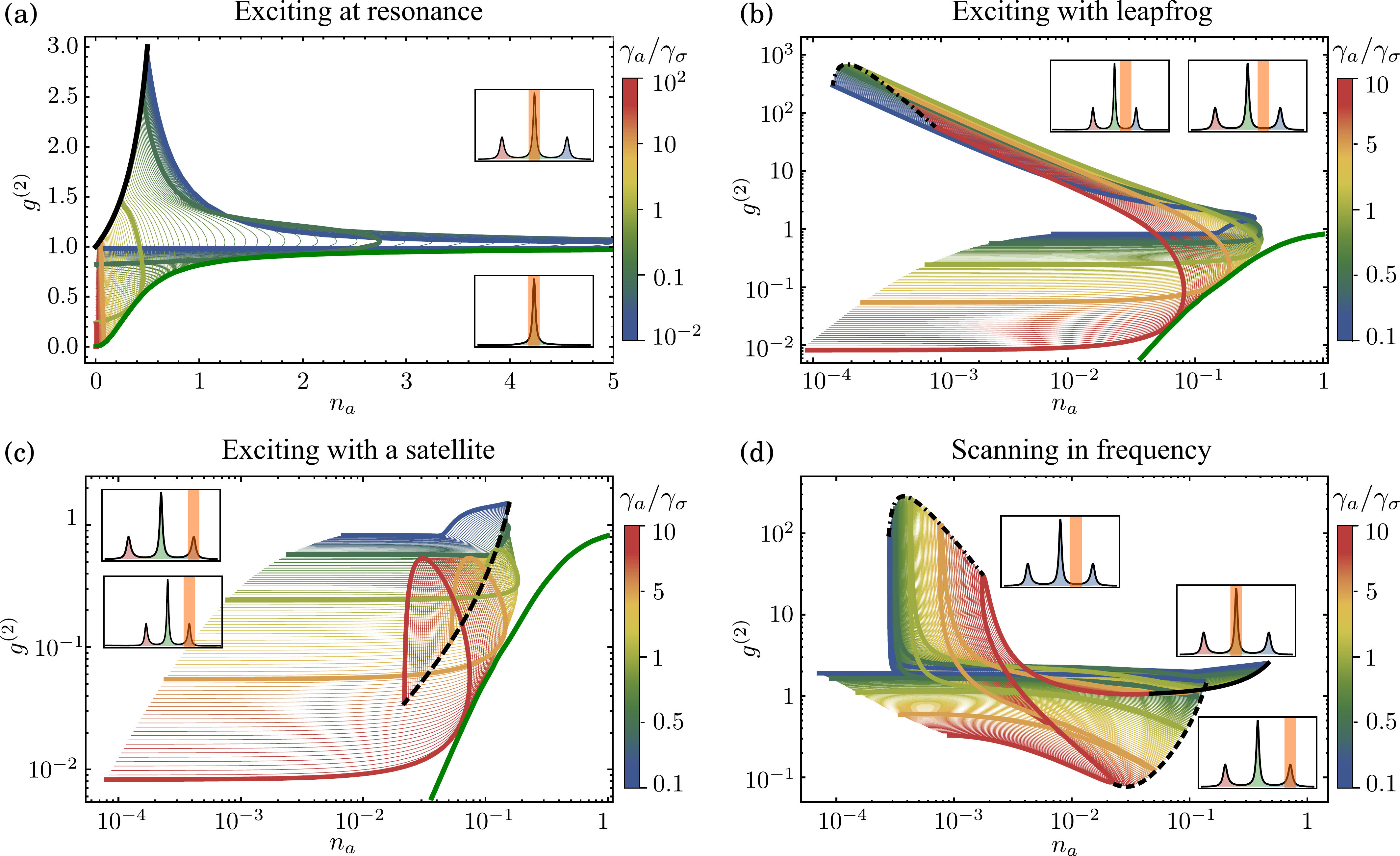}
    \caption{(Color online). Same as
      Fig.~\ref{fig:juedic17170852CET2015} but for the excitation from
      a coherent SPS. In (a,c) the implicit parameter is pumping and
      the green line shows the closest one can get from the Fock-duos
      limit from the case at resonance.  In~(d) the implicit parameter
      is the detuning between the source and its target. (a) Exciting
      at resonance. The optimum antibunching is the envelope in this
      configuration of the family of curves varying
      with~$\gamma_a/\gamma_\sigma$.  Highlighted are the
      cases~$\gamma_a/\gamma_\sigma=10^{-2}$ (blue), $10^{-1}$
      (green), 1 (chartreuse), 10 (yellow) and $10^2$ (red).  (b-d)
      Exciting in other configurations, namely (b) with the leapfrog
      processes, (c) with the satellite and (d) as a function of
      frequency. Highlighted are the
      cases~$\gamma_a/\gamma_\sigma=10^{-1}$ (blue), $0.5$
      (chartreuse), 1 (green), 5 (orange) and $10$ (red). The
      dashed-dotted line in panel (b) and the dashed line in panel (c)
      show the limit of large intensities and are merely a visual aid
      rather than a physical boundary. They are also shown in
      panel~(d), along with the black solid line for the envelope of
      the curves at $\omega_a=\omega_\sigma$. }
  \label{fig:domnov29120929CET2015}
\end{figure*}

\section{Exciting with a Coherent SPS}
\label{sec:marnov24163921CET2015}

We now consider excitation by a coherent SPS, i.e., when the 2LS that
acts as the source is driven coherently by an external laser, as
sketched in Fig.~\ref{fig:jueoct29162857CET2015}~d). This driving is
described by the Hamiltonian:
\begin{equation}
  H_\sigma = \omega_\sigma \ud{\sigma}\sigma + \Omega_\sigma \sqrt{\epsilon_1}
  (\ud{\sigma}e^{-i\omega_\mathrm{L} t} + \sigma e^{i\omega_\mathrm{L} t})\,,
\label{Eq.CoherentExcitationDot}
\end{equation} 
where $\omega_\sigma$ is the energy of the 2LS, and
$\omega_\mathrm{L}$ and
$\Omega_\sigma=\sqrt{\gamma_\sigma}\mathcal{E}$ are the frequency and
intensity of the driving laser (which drives a 2LS of decay rate
$\gamma_\sigma$ with a coherent field of intensity $\mathcal{E}$),
respectively. The coefficient $\sqrt{\epsilon_1}$ is put here so that
the cavity is driven only by the emission of the 2LS and not by the
external laser. The system is otherwise still described by the master
equations~(\ref{Eq.MasterEquationNormal})~and~(\ref{Eq.MasterEquationCoherent2}),
with $c_1=\sigma$ the 2LS (source) and $c_2=a$ the harmonic oscillator
(target), and reducing the coupling strength by a factor
$\sqrt{1-\epsilon_1}$, i.e., the coupling is given by
$\sqrt{(1-\epsilon_1)\gamma_a \gamma_\sigma}$. Here again, one can
obtain the cavity observables in closed-form, although it takes
slightly more cumbersome expressions. At resonance (out-of resonance
cases can also be obtained but get even more bulky):
\begin{widetext}
  \begin{subequations}
    \label{eq:juedic17152741CET2015}
    \begin{align}
      &n_a = \frac{16 (1-\epsilon_1)\gamma_{\widetilde{01}} \Omega_0^2 
        \left( \gamma_{\widetilde{11}}^2\gamma_{\widetilde{12}}+8\gamma_{\widetilde{10}} 
        \Omega_0^2\right)}{\gamma_{\widetilde{10}}\gamma_{\widetilde{11}}
        (\gamma_{\widetilde{01}}^2+8\Omega_0^2)(\gamma_{\widetilde{11}}
        \gamma_{\widetilde{12}}+16\Omega_0^2)}\,,\\
      &\nonumber\\
      &g^{(2)}=\nonumber\\
      &\frac{\gamma_{\widetilde{11}}(\gamma_{\widetilde{01}}^2+8\Omega_0^2)
        (\gamma_{\widetilde{11}}\gamma_{\widetilde{12}}+16\Omega_0^2)
        (\gamma_{\widetilde{11}}\gamma_{\widetilde{21}}^2\gamma_{\widetilde{31}}^2
        \gamma_{\widetilde{12}}\gamma_{\widetilde{32}}+8\gamma_{\widetilde{10}}
        \gamma_{\widetilde{31}}(17\gamma_{\widetilde{10}}^3+29\gamma_{\widetilde{10}}^2
        \gamma_{\widetilde{01}}+18\gamma_{\widetilde{10}}\gamma_{\widetilde{01}}^2+
        4\gamma_{\widetilde{01}}^3)\Omega_0^2+192\gamma_{\widetilde{10}}^2
        \gamma_{\widetilde{21}}\Omega_0^4)}{\gamma_{\widetilde{21}}
        \gamma_{\widetilde{31}}(\gamma_{\widetilde{11}}\gamma_{\widetilde{21}}+
        8\Omega_0^2)(\gamma_{\widetilde{31}}\gamma_{\widetilde{32}}+16\Omega_0^2)
        (\gamma_{\widetilde{11}}^2\gamma_{\widetilde{12}}+8\gamma_{\widetilde{10}}
        \Omega_0^2)^2}\,,
    \end{align}
  \end{subequations}
\end{widetext}
where we have introduced the notation
$\gamma_{\widetilde{ij}}=i \gamma_a + j\gamma_\sigma$ (e.g.,
$\gamma_{\widetilde{31}}=3\gamma_a+\gamma_\sigma$), and we have set
$\Omega_0 = \sqrt{\epsilon_1}\Omega$.  Eliminating~$\Omega^2$ from
Eqs.~(\ref{eq:juedic17152741CET2015}) yields, for a
given~$\gamma_a/\gamma_\sigma$, solutions of the type
$g^{(2)}=(P[3]\pm\sqrt{P[6]})/(P[2]n_a^2)$ where $P[n]$ is an $n$-th
order polynomial of the variable~$n_a$. The $\pm$ terms provide the
two branches of the curve as seen in
Fig.~\ref{fig:domnov29120929CET2015}(a). The lower branch corresponds
to low pumping, where the PL of the source is still a single line as
shown in the bottom inset, while the upper branch corresponds to high
pumping where the PL has split into a Mollow triplet, shown in the
upper inset.  Each curve starts from the point:
\begin{equation}
  \label{eq:vieene15153629CET2016}
  \left(n_a=0,g^{(2)}=\frac{1}{(1+\gamma_a/\gamma_\sigma)^2}\right)\,,
\end{equation}
showing that coherent SPS provide a much stronger antibunching than
their incoherent counterpart, Eq.~(\ref{eq:juedic17173145CET2015}),
already at vanishing pumping. At high pumping, in contrast to the
incoherent case that vanishes the population of its target, the
coherent SPS quenches it to a nonzero value and featuring bunching:
\begin{equation}
  \label{eq:vieene15155413CET2016}
  \left(n_a=\frac{1-\epsilon_1}{1+\gamma_a/\gamma_\sigma},g^{(2)}=3\frac{1+\gamma_a/\gamma_\sigma}{1+3\gamma_a/\gamma_\sigma}\right)\,.
\end{equation}
Eliminating~$\gamma_a/\gamma_\sigma$ here leads to the curve
$g^{(2)}=3(1-\epsilon_1)/[3(1-\epsilon_1)-2 n_a]$, which for
$\epsilon_1=1/2$ simplifies to $g^{(2)}=3/(3-4n_a)$ as shown in thick
black in Fig.~\ref{fig:domnov29120929CET2015}(a).  The expression for
the lower envelope of maximum antibunching, in blue, is too
complicated to be derived fully in closed-form but we can find simple
asymptotic expressions by series expansion of
Eqs.~(\ref{eq:juedic17152741CET2015}), namely,
$g^{(2)}\approx 5n_a^{2}$ when~$n_a\ll1$ and
$g^{(2)}\approx 1-1/(3(n_a+5))$ when~$n_a\gg1$.  The states thus get
close to the ideal Fock limit for large populations, approaching unity
as~$1/(3n_a)$ rather than~$1/n_a$. There is no limit in the population
that can be excited in the target by the coherent SPS, in stark
contrast to the incoherent SPS that is bounded by unity. Although the
accessible states in the~$(n_a,g^{(2)})$ space remain quite some
distance away from the ideal limit, the coherent SPS provides a much
better antibunching than its incoherent counterpart, as can be seen by
comparing Figs.~\ref{fig:juedic17170852CET2015}
and~\ref{fig:domnov29120929CET2015}. Namely, it is still antibunched
when population is unity (with a maximum antibunching slightly over
1/2) and we have alreay commented how it allows for arbitrary large
populations, that still feature antibunching. This is achieved by
exciting at resonance targets of very long lifetimes as compared to
the source.

This describes the resonant situation. Since the coherent SPS has a
rich spectral structure, it opens new configurations of excitation in
the Mollow triplet regime beyond the central peak, such as exciting
with a satellite peak (as shown in
Fig.~\ref{fig:domnov29120929CET2015}(c)) or with the leapfrog window
(Fig.~\ref{fig:domnov29120929CET2015}(b)). In these cases we show the
results in log-log plots as only small populations are within reach
(unlike the case of resonance) and, in the case of leapfrog
excitation, also a huge bunching can be imparted to the target. The
envelope of optimum antibunching as obtained in resonance is reported
in these panels (also in green) for comparison, showing that these
alternative schemes do not enhance antibunching.  The excitation with
leapfrog processes allows to access new regions of the Hilbert space
related to superbunching, so it is a configuration that presents its
own interest.  This is achieved at the price of small
populations. Exciting with a satellite peak also conquers new
territories in the $(n_a,g^{(2)})$ space not accessible through either
the central peak or leapfrog emission in a small region $(n_a\approx
0.6,g^{(2)}\lesssim2)$. Panel~(d) shows the trajectories when varying
the frequency of emission, where the cases just discussed---resonance,
leapfrog and satellites---appear as boundaries of the complete
picture, thus showing that these configurations already give access to
all the accessible states.

We can also compare in the parameter space the coherent SPS with its
Hamiltonian counterpart, as we did for the incoherent SPS. This is
shown in the bottom row of Fig.~\ref{fig1}, that is also usefully
compared with the upper row (incoherent SPS excitation). For the
coherent SPS as well, both cascaded and Hamiltonian coupling are
qualitatively similar in their statistical layout when the 2LS has a
decay term, but also differ notably from the topography of the
associated intensities. This is, again, clear on the $(n_a,g^{(2)})$
space, with the maximum antibunching that can be obtained for a given
cavity population with cascaded coupling (green line) and Hamiltonian
coupling (red line). The Hamiltonian coupling with no decay of the 2LS
also results in qualitative differences and an enhanced
antibunching. However, in this case, the Hamiltonian coupling never
comes to surpass the cascaded scheme.  Instead, in all the range of
population, the antibunching obtained through the cascaded coupling is
larger than the one obtained with any type of normal (Hamiltonian)
coupling, for a given population. When the cavity is detuned from the
2LS, the antibunching is lost before the cavity's population reaches
one photon. Also, when the cavity is in resonance with the leapfrog
process emission, the photon statistics of the cavity is superbunched
but its population remains well below unity. Including a dephasing
rate to the dynamics of the 2LS increases the broadening of each
emission line in the triplet, and even a small dephasing rate spoils
the antibunching. Thus, the best antibunching with a coherent SPS is
obtained exciting the cavity at low pumping and in resonance with the
2LS, using the cascaded coupling.

\begin{figure}
    \includegraphics[width=0.9\linewidth]{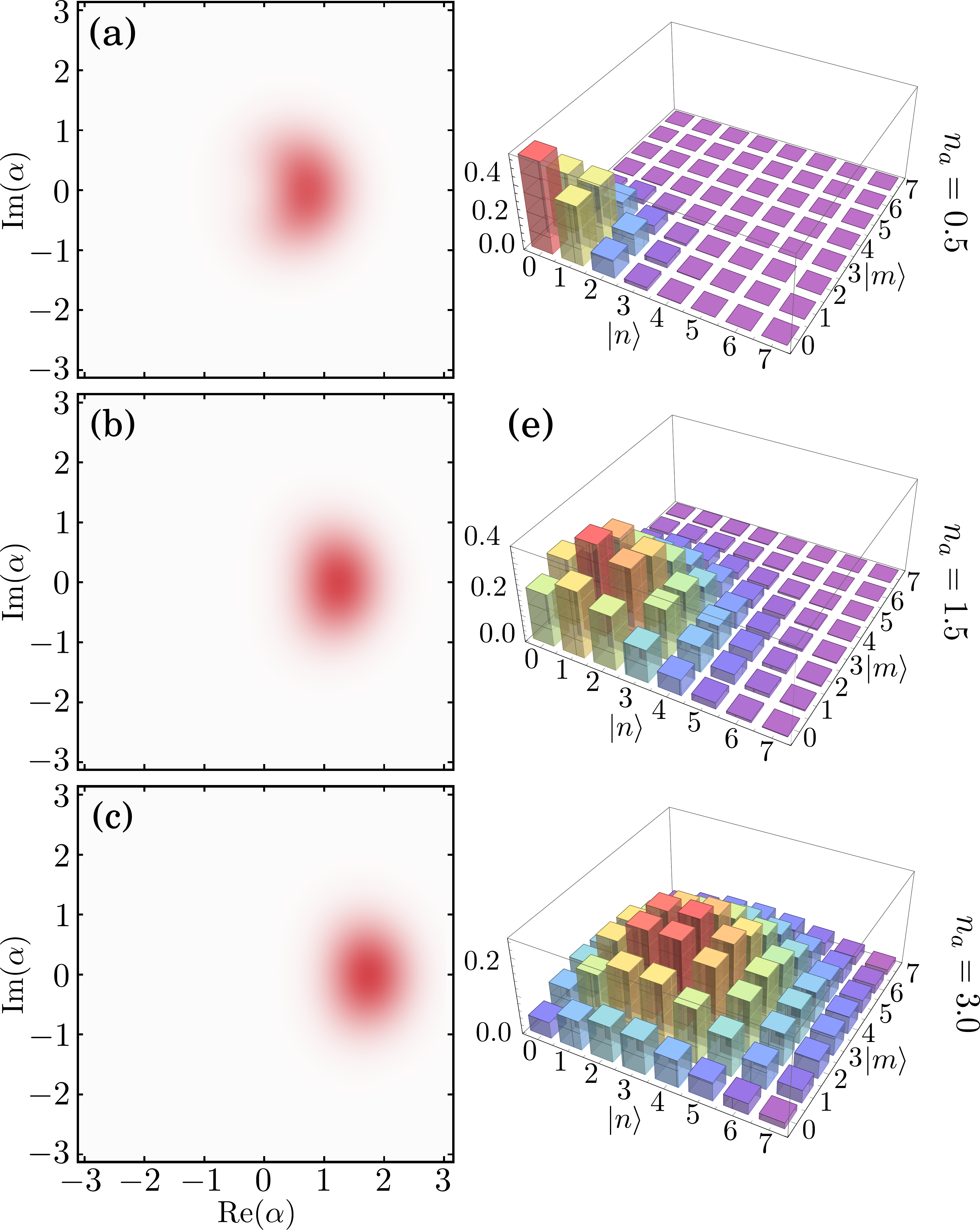}
    \caption{(Color online). Wigner function~(a--c)~and density
      matrix~(d--f) for the states marked by black circles along the
      green line in Fig.~\ref{fig1}(h). These states have
      populations~$n_a=1$, $1.5$ and~$3$ from top to bottom and all
      feature antibunching.}
  \label{fig4}
\end{figure}

One of the departing features of the coherent SPS excitation, is that
it allows to populate its target with more than one excitation on
average, therefore triggering Bose stimulation effects. There is a
phase locking (of $0$, as determined by the Hamiltonian) of the
incident photons from the quantum source that accumulate while still
exhibiting some features of Fock states, in particular being
antibunched. A complete picture of the resulting quantum states,
beyond the diagonal elements only, is given Fig.~\ref{fig4} for the
states marked with black circles in Fig.~\ref{fig1}(f), both through
the Wigner representation, Fig.~\ref{fig4}(a-c), and through their
matrix representation, Fig.~\ref{fig4}(d-f). The Wigner representation
exhibits negative values at the origin, a mark of a genuine quantum
state with no classical counterpart, for the states with $n_a=1$ and
$n_a=1.5$ (the blue spot at the center in Fig.~\ref{fig4}(b)). It is,
however, positive everywhere for~$n_a=3$, although the state is still
antibunched.  Also, one can see that the phase uncertainty decreases
as the population increases, in agreement with the ``classical
limit''. Moreover, the matrix elements in these cases also satisfy the
relation $\rho_{22} > \max(\rho_{11}/2,\rho_{33})$, which means that
fluctuations are smaller than could be expected from rate equation
arguments, since two excitations are twice as much likely to decay
than one excitation, making the probability to find one excitation
only in principle at least twice as large than double
occupation.\cite{lopezcarreno15a}

\begin{figure}
  \includegraphics[width=0.8\linewidth]{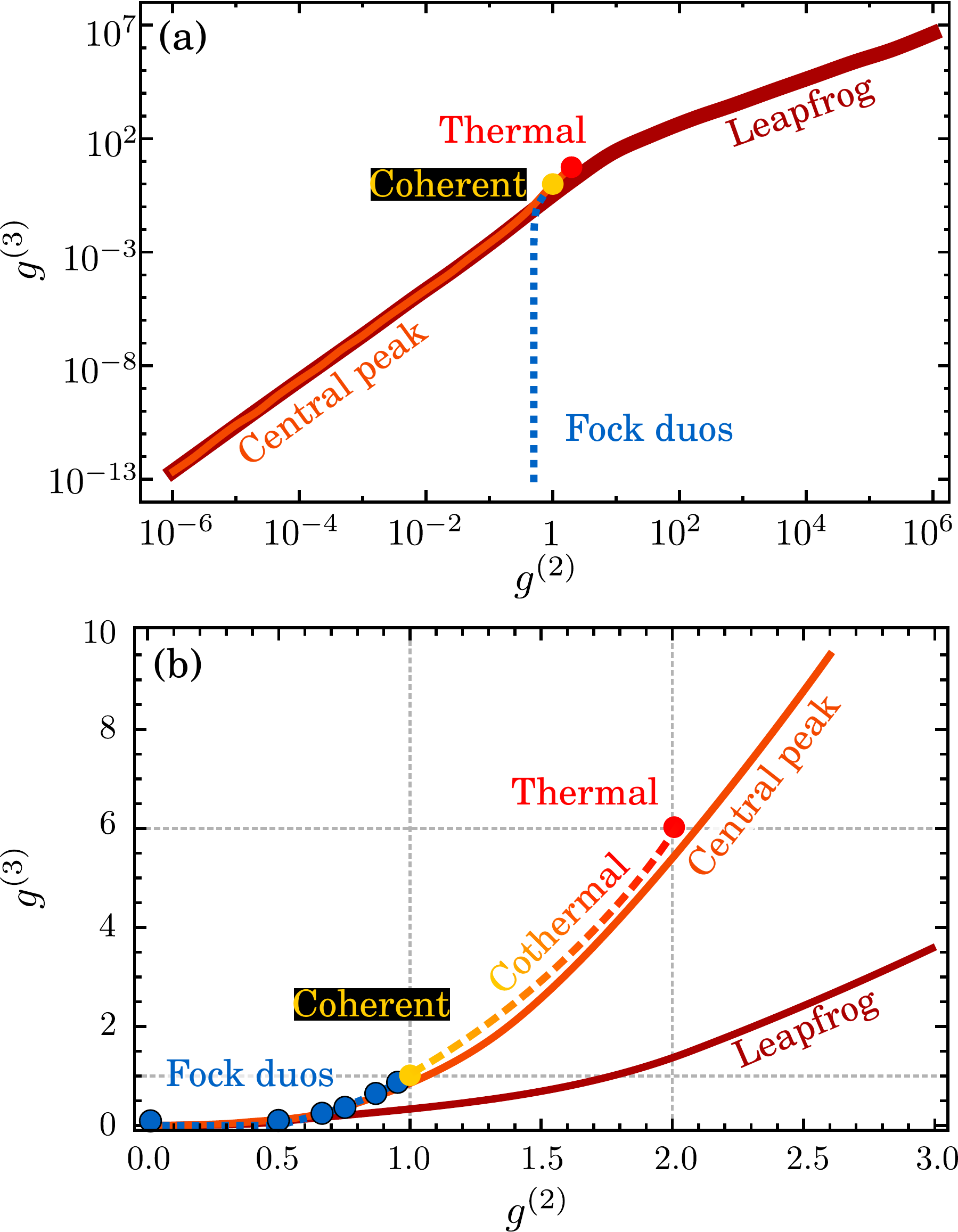}
  \caption{(Color online).  States accessible in the
    $(g^{(2)},g^{(3)})$ space by coherent SPS excitation, at resonance
    with the central peak (orange line) or with the leapfrog emission
    (red line). The dashed blue line corresponds to the ``Fock duos'',
    Eq.~(\ref{eq:vieoct23174124CEST2015}). The same result is shown in
    both (a) log-log and~(b) linear scales.}
  \label{fig:vieoct23155212CEST2015}
\end{figure}

We conclude this Section on driving an harmonic oscillator with the
coherent SPS by considering states in the $(g^{(2)},g^{(3)})$
space. This is shown in Fig.~\ref{fig:vieoct23155212CEST2015} for the
most notable configurations of exciting with the central peak of the
Mollow triplet or the leapfrog region between the central peak and a
satellite. All points are physical on this map, including cases with
$g^{(2)}\ll 1$ and $g^{(3)}\gg 1$ (and vice-versa), although the
accessible regions are strongly confined along curves that we can fit
in monomial form, leading in good approximation to~$g^{(3)}\approx 0.2
[g^{(2)}]^2$ at low pumping (in the antibunching corner)
and~$g^{(3)}\approx 4.5 g^{(2)}$ at large pumpings (in the
superbunching corner). We indicate positions of popular quantum
states such as Fock states, coherent states, thermal states and their
combinations. Driving with a SPS spans over great distances in this
space, especially when exciting with leapfrog processes, as seen in
panel~(a) of the figure. This shows, again, how the central peak
remains confined essentially to the antibunched corner of the map
(excursions to bunching cases are up to~$g^{(2)}=3$). Panel~(b) is a
zoom around small values in linear scale.  The leapfrog processes, on
the contrary, allow one to reach large joint values of $g^{(2)}$
and~$g^{(3)}$. Interestingly, while the points accessible in
the~$(n_a,g^{(2)})$ space extend over 2D~areas, in the
$(g^{(2)},g^{(3)})$ space they are confined to 1D~curves, showing how
these correlators are strongly inter-dependent, and thus making
the~$(n_a,g^{(2)})$ representation more prominent.

\section{Variations in the type of coupling}
\label{sec:lunnov30102926CET2015}

The comparison between both rows of Fig.~\ref{fig1} shows that the
coherent SPS is a much better quantum drive to generate bright
antibunching than the incoherent one, even in the region where
$n_a<1$.  The coherent SPS exciting the cavity still remains some
distance away from the Fock duos boundary, but it gets significantly
closer than with the Hamiltonian coupling.  We complete our
juxtaposition between the two types of coupling, cascaded and
Hamiltonian, by comparing with other variations in the way the source
couples to its target.

\begin{figure}
  \includegraphics[width=0.95\linewidth]{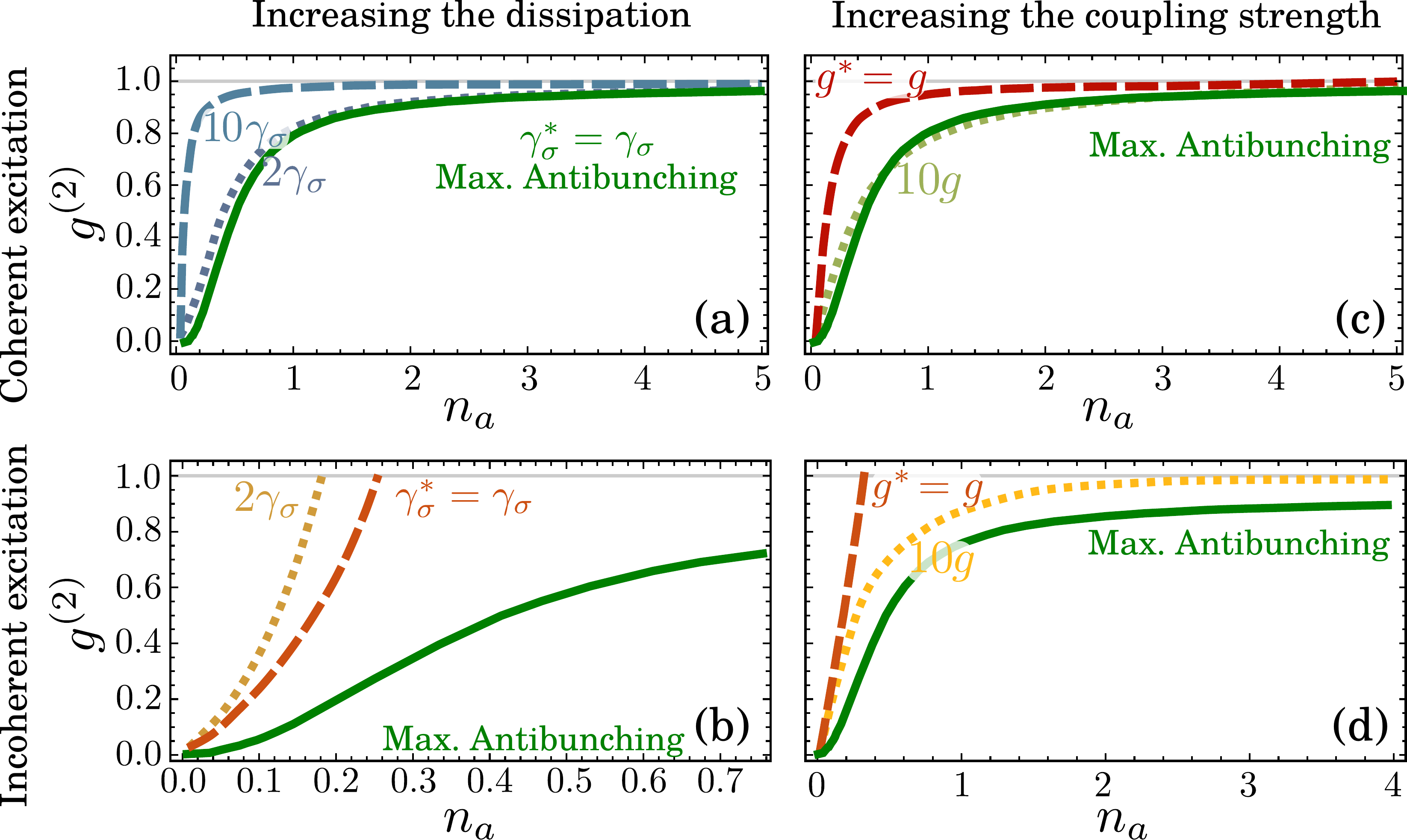}
  \caption{(Color online). Variations in the model. (a,~b)~Adding a
    dissipative channel to the source in the cascaded coupling spoils
    the antibunching. (c,~d)~Increasing the coupling strength $g$ in
    the Hamiltonian coupling (with decay in the source) enhances the
    antibunching until it reaches the one obtained for the Hamiltonian
    coupling without decay of the source (when $g^\ast \approx
    10g$). However, the enhancement is never enough to top the
    antibunching obtained using the coherently driven 2LS and the
    cascaded coupling.}
  \label{fig2}
\end{figure}

First, we consider another source of decay for the source in the
cascaded coupling scheme, i.e., with an extra term
$(\gamma_\sigma^\ast/2)\mathcal{L}_{\sigma}\rho$ in the master
equation, leaving the coupling strength constant and equal to
$\sqrt{\gamma_a \gamma_\sigma}/2$. This describes the situation where
not all the light that is lost from the source is redirected to the
target. Figure~\ref{fig2}(a-b) shows how this other source of decay
spoils the antibunching and drives the cavity faster toward a coherent
state (for the coherent SPS) or to a thermal state (for the incoherent
SPS). At $\gamma_\sigma^\ast=10\gamma_\sigma$, the antibunching is
worse than that obtained using the Hamiltonian coupling (without any
additional dissipative channel). This shows that it is better to
minimize losses from the quantum emitter, but that unless these are
drastic, the impact is moderately detrimental.  Note that the cascaded
coupling strength can only be smaller or at most equal to
$\sqrt{\gamma_1\gamma_2}$, unlike some statements in the
literature~\cite{flayac14a} that refer to ``arbitrary coupling
strengths''. This leads, otherwise, to unphysical states for the
target.  Second, in the Hamiltonian coupling, we increase the coupling
strength and bring the system into the strong-coupling regime. This is
one considerable degree of freedom of the Hamiltonian coupling that is
lost when considering some external excitation, since photons that
drive the system from outside can couple to it in the first place
because it is dissipative. In fact, the more efficient is the
source--target coupling, the more dissipative the target has to
be. But a dissipative target cannot sustain quantum effects for long,
as the states do indeed decay. Therefore there is a compromise for the
target's lifetime to allow simultaneously an efficient driving from
the source while still allowing to store long enough the imprinted
quantum attributes. In the Hamiltonian coupling, however, coupling and
decay are independent and one can consider a pure Hamiltonian picture
with no decay whatsoever. We will see later in this series how one can
remedy to this vicious circle to some extent by considering other
types of quantum sources. For now, we show that, surprisingly,
cascaded coupling indeed provides a better quantum driving than the
Hamiltonian coupling brought in the strong-coupling regime. This is
because in the latter case, the reversible process from the ``target''
back to the ``source'' acts like an effective decay. This is shown in
Fig.~\ref{fig2}(c,d), where we compare again the maximum antibunching
obtained in the configuration of Sec.~\ref{sec:marnov24163921CET2015}
(solid green lines) with a magnified Hamiltonian coupling
$g^\ast = N \sqrt{\gamma_a \gamma_\sigma}/2$ for various $N$. For both
the coherent and incoherent excitation cases, the antibunching is
improved by raising the coupling strength. However, this improvement
is still not large enough to overcome the cascaded coupling. As seen
in the figure, even for large Hamiltonian coupling strengths, the
antibunching obtained with the cascaded coupling (solid green line) is
always significantly better.  Increasing even further the value of $N$
does not improve the antibunching, but makes it occur at larger
$(\Omega_\sigma,\gamma_a)$ values. At this limiting case, the
antibunching obtained for the Hamiltonian coupling with the decay of
the source coincides with the antibunching for the Hamiltonian without
the decay of the source.  Therefore, Fid.~\ref{fig2}(c) shows the
optimum antibunching that can be obtained using both the cascaded and
Hamiltonian couplings.

Increasing the coupling has a much stronger effect in the case of
incoherent excitation, as shown in Fig.~\ref{fig2}(d). Even a small
variation of the coupling strength, $N=1.5$, changes significantly the
curvature of the antibunching curve. Increasing the value of $N$ makes
the system enter the ``one-atom laser'' regime,\cite{delvalle11a} and
the state in the cavity becomes coherent. As for the coherent
excitation, the best antibunching saturates with increasing coupling
(converging at roughly $g^\ast=10g$), with larger values of the
coupling strength merely occuring at larger $(P_\sigma,\gamma_a)$
values. Nevertheless, although raising the coupling strength improves
the antibunching, and at large $g^\ast$ the $g^{(2)}$ of the cavity at
large population goes to $1$, the best antibunching is still obtained
with the cascaded coupling in the coherent excitation regime (green
line in Fig.~\ref{fig2}(d)). Overall, it is therefore established that
optimum antibunching is obtained in the cascaded architecture. This
could have important consequences in the design of future
quantum-optical devices.

\section{Summary and Conclusions}
\label{sec:jueoct29112554CET2015}

\begin{figure}
  \includegraphics[width=0.95\linewidth]{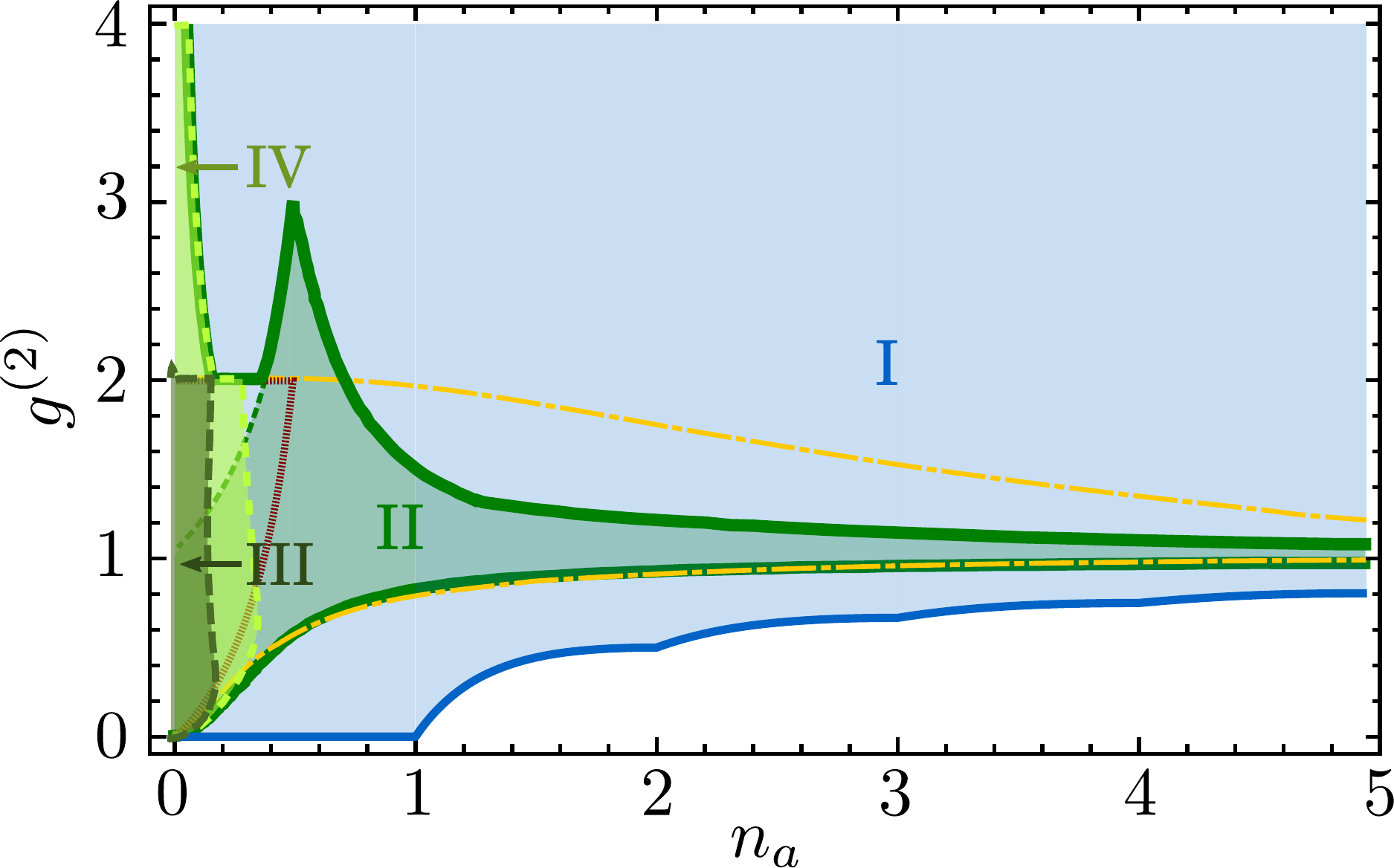}
  \caption{(Color online). Summary of our results: the Hilbert space
    of an harmonic oscillator charted by the~$n_a$ (population)
    and~$g^{(2)}$ (second-order correlation function) is accessible
    for any point in the blue region~I (the white region is
    forbidden). Of this region, the wider span is achieved by exciting
    with a coherent SPS, covering the green region. This region breaks
    down into~II, the region achieved at resonance between the source
    and the target, III, that achieved with the excitation from a
    satellite of the Mollow triplet and~IV that achieved with the
    leapfrog emission. The incoherent SPS covers a smaller region,
    deliminited by the red dotted curve. The hamiltonian coupling,
    between the two dotted-dash yellow curves, also covers a large
    span of the Hilbert space and is neither included nor fully
    encludes the SPS area.  In particular, the excitation by a SPS
    allows to provide a stronger antibunching for any given population
    as compared to a Hamiltonian coupling, although it still remains
    some distance away from the ideal limit.}
  \label{fig:miéene13175933CET2016}
\end{figure}

Our analysis of the excitation by a SPS of a quantum harmonic
oscillator can be summarized by Fig.~\ref{fig:miéene13175933CET2016}.
Here we show again the charting of the Hilbert space of the quantum
harmonic oscillator that we have introduced, along with the various
areas that can be accessed with a SPS. First, the region~I of physical
states, shown in light blue, delimited by Fock duos
(Eq.~(\ref{eq:marnov24092156CET2015})) according to
Eq.~(\ref{eq:miéoct21181551CEST2015}). No state exists that can
provide the corresponding joint population and antibunching in the
white region.  In contrast to popular belief, states do exist in the
region above this boundary that satisfy~$g^{(2)}<1/2$ and $n_a>1$
which shows that on mathematical grounds, the criterion $g^{(2)}<1/2$
cannot be used to exclude states with more than one particle. In this
region of physical states, the green area can be accessed in the
steady state by exciting a cavity with a SPS, provided an
adequate~$\gamma_a/\gamma_\sigma$ ratio and pumping. The largest span
is realized by a coherent SPS (i.e., a 2LS itself driven by a laser)
at resonance with its target. This gives access to the light green
region~II. Not all states are accesible in this configuration, in
particular the limit of Fock duos remains some distance away from the
driven dissipative case. We will show in Part~II of this work how one
can get closer using other quantum sources. States of arbitrary
populations can be reached with the coherent SPS (by using targets
with long enough lifetimes) that still retain antibunching. One can
also realize states of arbitrary high bunching, by using leapfrog
emission of the SPS in the Mollow triplet regime. This is the
region~IV shown in dark green in
Fig.~\ref{fig:miéene13175933CET2016}. Exciting with the satellite
peaks covers the region~III that gives access to a small patch not
within reach of~II and~IV. Other frequencies do not lead to areas not
already covered by these cases. The incoherent SPS does not extend
this territory either, as it is contained in the
area~$2n_a/(3-2n_a)\le g^{(2)}\le2$, shown in dotted red.  Notably,
the region accessible by the Hamiltonian regime does not include nor
is included by that accessible with the cascaded coupling.  They do
share some common area.  In fact, cascaded coupling reduces to
Hamiltonian coupling whenever
$g/\gamma_\sigma\ll\gamma_a/\gamma_\sigma$ which, for the cascaded
coupling constrain of
$g/\gamma_\sigma=\sqrt{\gamma_a/\gamma_\sigma}/2$, provides the
criterion~${\gamma_a}/{\gamma_\sigma}\gg1/4$ for which the
conventional formalism is enough to describe excitation with no
feedback. This criterion allows, for instance, to develop the theory
of frequency correlation~\cite{delvalle12a} in a conventional setting
without the explicit need of the cascaded formalism to prevent
feedback, a point that has not always been fully
appreciated.\cite{flayac14a} In other cases, we have shown that
cascaded coupling allows, surprisingly, to reach regions of the
Hilbert space out of reach of a conventional Hamiltonian dynamics and,
in particular, to improve the possible antibunching for a given
population.

\begin{figure}
    \includegraphics[width=.85\linewidth]{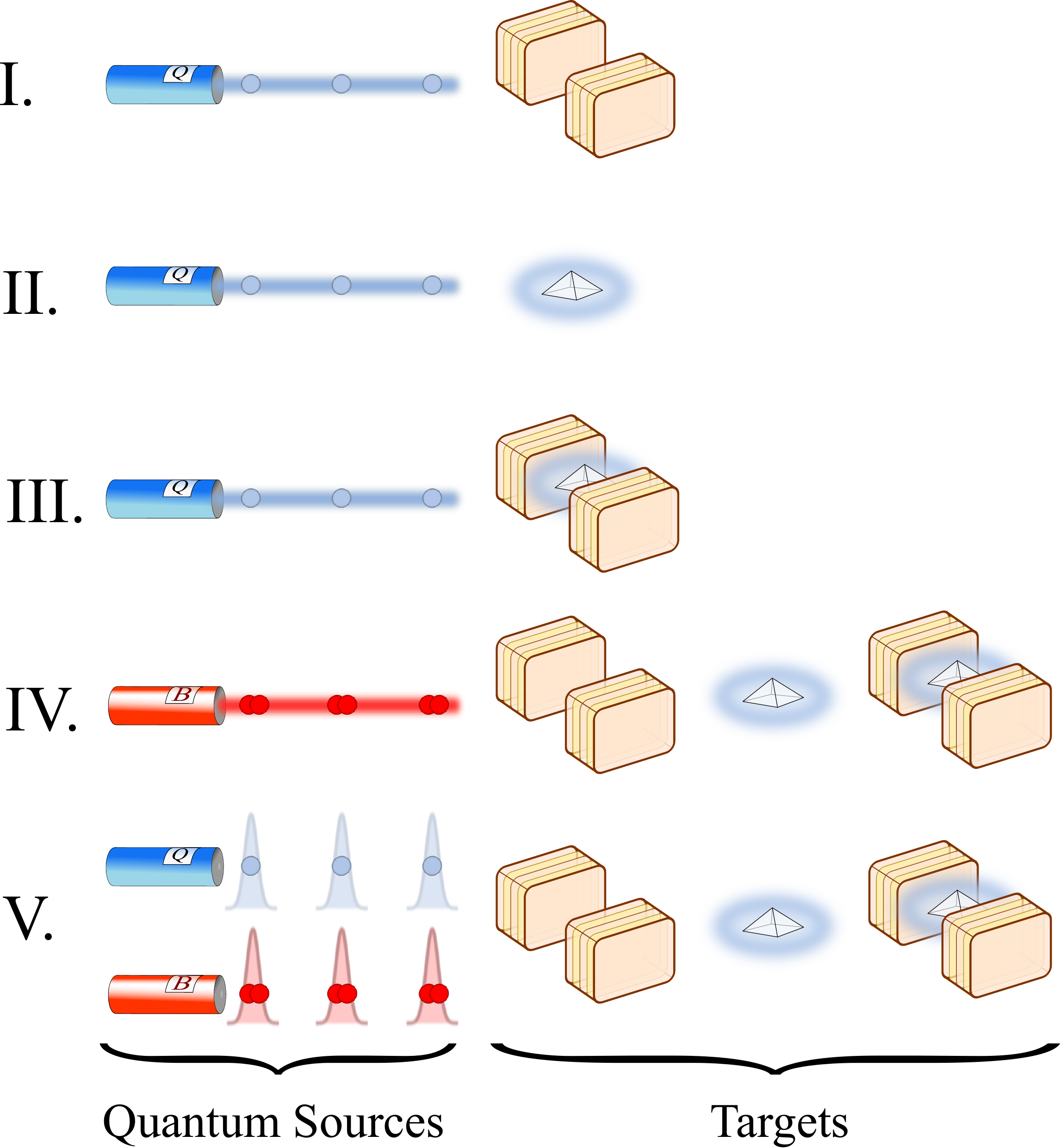}
    \caption{(Color online). Exciting with Quantum Light: the basic
      configurations we propose to study in this series of
      texts. Part~I is the current work, exciting an harmonic
      oscillator (single cavity mode) with a SPS. Part~II addresses
      the excitation of a 2LS with a SPS. Part~III is exciting coupled
      light-matter systems, or polaritons, still with a SPS. Following
      the comprehensive treatments of these several targets under the
      cw excitation from a SPS, we then trade this simple quantum
      source for, IV, a $N$--photon emitter (bundler) and, V, a pulsed
      quantum source, impinging on all the targets previously
      considered.}
  \label{fig:vieoct30160938CET2015}
\end{figure}

\section{Perspectives}
\label{sec:jueoct29112601CET2015}

We have considered only one aspect of the general picture of quantum
excitation of optical targets. Namely, various types of SPS exciting
an harmonic oscillator. In the laboratory, this can take the form of
exciting a passive cavity mode, or a non-interacting boson field. This
is the configuration sketched in
Fig.~\ref{fig:vieoct30160938CET2015}~I.~and that we addressed in this
text, the opening one of a series that will carry on similar studies
to other configurations of interest, as sketched in the rest of
Fig.~\ref{fig:vieoct30160938CET2015} and summarized here.  In the next
text, II., we consider the problem of exciting a 2LS rather than the
harmonic oscillator. In part~III, we increase one step more the level
of complexity of the target by exciting polaritons, i.e., coupled
light--matter systems. This focuses on different targets subjected to
the (quantum) light of the same type of source. In part~IV, we revisit
all these targets together but now excited by an even ``more quantum''
source than the SPS, namely, $N$-photon
emitters.\cite{sanchezmunoz14a} In Section~V., we come back to all the
combinations dealt with so far but turning to pulsed excitation rather
than cw, addressing the problem of quantum state preparation. Even
with the five texts taken together, we do not exhaust the topic and
much remains to be studied even in the configurations put to
scrutiny. For instance, we considered one particular type of SPS only,
the simplest one as realized by a single emitter, while
implementations from pairs of photons with one heralding the other
(mainly through parametric down conversion~(PDC) and four-wave mixing
(FWM)) are extremely popular and with advantages of their own, such as
determinism. It is clear that there is much engineering to be done
with such sources as well, and certainly some improvements in their
operating conditions to be gained from the cascaded architecture,
e.g., helping to fix their principal drawback so far: the nonzero
probability to emit more than a single photon.  Also, this extended
exploration of the general problem of quantum excitation will give us
many occasions to revisit the same aspects from different
angles. Already in part~II, we will come back to the question of what
regions of the Hilbert space are within reach. As another example, in
part~V, we will provide a clear physical picture of why the cascaded
architecture allows such an enhancement as compared to the Hamiltonian
coupling by considering the time-resolved excitation of the target by
the spontaneous emission of an initial condition prepared in the
source, which can be solved exactly. This brings much clarification
into the nature of the target and the source and their
interrelationship in a quantum context.

\begin{acknowledgments}
  We thank C. S\'anchez Mu\~noz and E. del Valle for constant
  discussions.  Funding by the POLAFLOW ERC project No.~308136 and by
  the Spanish MINECO under contract FIS2015-64951-R (CLAQUE) is
  acknowledged.
\end{acknowledgments}
\appendix
\section{Proof that Fock duos bound the $(n_a,g^{(2)})$ space}
\label{sec:marnov24122827CET2015}

We prove mathematically that there are no states below the boundary
set by Eq.~(\ref{eq:miéoct21181551CEST2015}) but at least one for
every point above, thus showing that our charting of the Hilbert space
in the $(n_a,g^{(2)})$ space as defined by this boundary is
complete.\\

\begin{proposition}
  \emph{For a given population $n_a\in\mathbb{R}$, the maximum
    antibunching is given by a superposition of at most two
    consecutive Fock states.}
\end{proposition}

First, for any~$n_a\in\mathbb{R}$, there exists a superposition of at most
two consecutive Fock states that provides this population, namely:
\begin{equation}
  \label{eq:vieoct23174124CEST2015}
  \sqrt{\lfloor n_a \rfloor-n_a+1}\ket{\lfloor n_a \rfloor}+\sqrt{n_a-\lfloor n_a \rfloor}\ket{\lfloor n_a \rfloor+1}\,.
\end{equation}
We can write without loss of generality the generic
state~(\ref{eq:miéoct21094346CEST2015}) as:
\begin{equation}
  \label{eq:vieoct23175317CEST2015}
 \sum_{k=-\lfloor n_a \rfloor}^\infty\sqrt{p(\lfloor n_a \rfloor+k)}\ket{\lfloor n_a \rfloor+k}\,,
\end{equation}
where we remind that $p(n)=|\alpha_n|^2$ and is such that
$\sum_{n=0}^\infty p(n)=1$. Restricting ourselves to states with the
same population implies:
\begin{equation}
  \label{eq:vieoct23181019CEST2015}
  \sum_{k=-\lfloor n_a \rfloor}^\infty p({\lfloor n_a \rfloor+k})(\lfloor n_a \rfloor+k)=n_a\,.
\end{equation}
The second order correlation function for such states is then found as:
\begin{equation}
  \label{eq:vieoct23181207CEST2015}
  g^{(2)}=\sum_{k=-\lfloor n_a \rfloor}^\infty (\lfloor n_a \rfloor+k)(\lfloor n_a \rfloor+k-1)p(\lfloor n_a \rfloor+k)/n_a^2\,.
\end{equation}

Since $n_a$ is the same for both
states~(\ref{eq:vieoct23174124CEST2015})
and~(\ref{eq:vieoct23175317CEST2015}) when
Eq.~(\ref{eq:vieoct23181019CEST2015}) is satisfied, it is enough for
the comparison of their antibunching to consider the difference of
their $n_a^2g^{(2)}$ (effectively getting rid of the denominator,
which simplifies the notations). We call this difference~$\Delta
G^{(2)}$ and find, from Eq.~(\ref{eq:miéoct21181551CEST2015})
and~(\ref{eq:vieoct23181207CEST2015}):
\begin{widetext}
  \begin{align*}
    \Delta G^{(2)} &= \sum_{k=-\lfloor n_a \rfloor}^\infty (\lfloor n_a \rfloor+k)(\lfloor n_a \rfloor+k-1)p({\lfloor n_a \rfloor+k}) -\lfloor n_a \rfloor(\lfloor n_a \rfloor-1)(\lfloor n_a \rfloor-n_a+1)-\lfloor n_a \rfloor(\lfloor n_a \rfloor+1)(n_a-\lfloor n_a \rfloor)+{}\,,\\
    &
    \begin{aligned}
      {}=&\left[\sum_{k\neq \lbrace 0,1\rbrace}(\lfloor n_a
        \rfloor+k)(\lfloor n_a \rfloor+k-1)p({\lfloor n_a
          \rfloor+k})\right]+{}\\&{}+\lfloor n_a \rfloor(\lfloor n_a
      \rfloor-1)(p({\lfloor n_a \rfloor})-\lfloor n_a
      \rfloor+n_a)+\lfloor n_a \rfloor(\lfloor n_a
      \rfloor+1)(p({\lfloor n_a \rfloor+1})-(n_a-\lfloor n_a
      \rfloor))\, ,
    \end{aligned}
  \end{align*}
\end{widetext}
where we have separated the terms $k=0,1$ from the sum and re-arranged
the factors. Now, since $p(\lfloor n_a \rfloor+1) = 1-p(\lfloor n_a
\rfloor) -\sum_{k\neq \lbrace 0,1\rbrace} p({\lfloor n_a \rfloor+k})$
from the probability normalization, we arrive to:
\begin{multline}
  \label{eq:vieoct23190530CEST2015}
  \Delta G^{(2)}=
  \left[\sum_{k\neq \lbrace 0,1\rbrace} (k-1)(2\lfloor n_a \rfloor+k)p({\lfloor n_a \rfloor+k})\right]\\{}-2\lfloor n_a \rfloor (p({\lfloor n_a \rfloor})-{\lfloor n_a \rfloor}+n_a-1)\,.
\end{multline}
On the other hand, from Eq.~(\ref{eq:vieoct23181019CEST2015}), we can isolate:
\begin{equation}
 p({\lfloor n_a \rfloor}) = \lfloor n_a \rfloor-n_a+1+\sum_{k\neq \lbrace 0,1\rbrace }(k-1)p({\lfloor n_a \rfloor+k})\,,\\
\end{equation}
which, injected back in Eq.~(\ref{eq:vieoct23190530CEST2015}), yields:
\begin{align*}
  \Delta G^{(2)}=\sum_{k\neq \lbrace 0,1\rbrace} k(k-1)p({\lfloor n_a \rfloor+k})\, .\\
\end{align*}
This is the final result, since all the terms in the sum are
non-negative, we can conlude that~$\Delta G^{(2)}\ge 0$. The
inequality is maximized when \emph{all} the probabilities $p({\lfloor
  n_a \rfloor+k})=0$ for $k\neq\lbrace0,1\rbrace$, i.e., for the
states of type~(\ref{eq:vieoct23174124CEST2015}).  When $n_a$ is an
integer, i.e., $\lfloor n_a \rfloor=n_a$, the maximum antibunching is
obtained with the Fock state $\ket{n_a}$. QED.

\begin{proposition}
  \emph{For a given population $n_a\in\mathbb{R}$ and any positive
    real number~$\delta$, there is at least one state such that
    $\Delta G^{(2)}=\delta$.}
\end{proposition}

Let us consider the set of states with population $n_a=\sum_n n p(n)$. We can 
construct a state $\rho_0$ with population $n_a$ in the following way:
\begin{multline*}
\rho_0 = p(0) \rhoel{0} +p(\lfloor n_a \rfloor) \rhoel{\lfloor n_a \rfloor} \\
+ (1-p(0) - p(\lfloor n_a \rfloor))\rhoel{k}\, .
\end{multline*}
Imposing the population constraint, we obtain the following condition:
\begin{equation*}
p(0) = \left( \frac{k-n_a}{k}\right)-p(\lfloor n_a \rfloor)\left(\frac{k-\lfloor n_a \rfloor}{k} \right)\, ,
\end{equation*}
where we have assumed that $\lfloor n_a \rfloor < k \in \mathbb{N}$. The $g^{(2)}=g^{(2)}_\ast+\delta$ of this state is given by:
\begin{multline*}
g^{(2)} n_a^2 = \lfloor n_a \rfloor(\lfloor n_a \rfloor-1)p(\lfloor n_a \rfloor) \\
+ k(k-1)(1-p(0)-p(\lfloor n_a \rfloor))\, ,
\end{multline*}
from which we can obtain $p(\lfloor n_a \rfloor)$:
\begin{equation*}
p(\lfloor n_a \rfloor) = \frac{n_a\left( k-1-n_a g^{(2)}\right)}{\lfloor n_a \rfloor(k-\lfloor n_a \rfloor)}\, .
\end{equation*}
If $\delta>0$ we can find an integer number $k> \textrm{max}(n_a
g^{(2)},\lfloor n_a \rfloor)$ such that $0\leq p(\lfloor n_a
\rfloor)\leq 1$. Considering that this state is not made of two
consecutive Fock states, the accessible $g^{(2)}$s are larger than
$g^{(2)}_\ast$. Of course, when $p(0)=0$ and $k=\lfloor n_a
\rfloor+1$, $\rho_0$ is the maximally antibunched state and $g^{(2)} =
g^{(2)}_\ast$.  When $n_a<1$, $\lfloor n_a \rfloor=0$, and the
procedure above fails. In such case, we can build another state,
\begin{equation*}
\rho_1 = p(0) \rhoel{0} + p(1) \rhoel{1} + (1-p(0)-p(1))\rhoel{k}\, ,
\end{equation*}
such that $n_a = p(1) + k(1-p(0)-p(1))$. From this constraint we
obtain the following condition:
\begin{equation*}
p(1) = \frac{k(1-p(0))-n_a}{k-1}\, .
\end{equation*}
The $g^{(2)}$ of this state is given by:
\begin{align*}
g^{(2)}n_a^2 &= k(k-1)(1-p(0)-p(1))\, ,\\
	&= k(n_a+p(0)-1)\, ,
\end{align*}
from which we conclude that:
 \begin{equation*}
 p(0) = \frac{g^{(2)}n_a^2+k(1-n_a)}{k}\,.  
 \end{equation*}
Again, we can always find an integer $k>1$, such that $0\leq p(0) \leq 1$. 

\bibliographystyle{naturemag}
\bibliography{Sci,arXiv,books}

\end{document}